\documentclass[apj]{emulateapj}

\newcommand{\etal }{{et al.} }
\newcommand{\msun}{\thinspace M_\odot}

\newcommand{\vect}[1]{\mbox{\boldmath$#1$}}
\def\lesssim{\mathrel{\hbox{\rlap{\hbox{\lower4pt\hbox{$\sim$}}}\hbox{$<$}}}}
\def\gtrsim{\mathrel{\hbox{\rlap{\hbox{\lower4pt\hbox{$\sim$}}}\hbox{$>$}}}}
\newcommand{\cm}{\,{\rm cm}^{-3} }

\newcommand{\rcri}{R_{\rm c} }

\newcommand{\dfrac}[2]{{\displaystyle \frac{#1}{#2}} }
\newcommand{\ang}{\delta_0}

\usepackage{color}

\shorttitle{Magnetic Field Structure in Star-Forming Cores}
\shortauthors{Kataoka \etal 2012}

\begin{document}

\title{Exploring Magnetic Field Structure in Star-Forming Cores with Polarization of Thermal Dust Emission}
\author{Akimasa Kataoka\altaffilmark{1,2,3}, Masahiro N. Machida\altaffilmark{4}, and Kohji Tomisaka\altaffilmark{2,3}} 
\altaffiltext{1}{Department of Astronomy, Kyoto University, Sakyo-ku, Kyoto 606-8502, Japan}
\altaffiltext{2}{Department of Astronomical Science, School of Physical Sciences, Graduate University for Advanced Studies (SOKENDAI), Osawa 2-21-1, Mitaka, Tokyo 181-8588, Japan}
\altaffiltext{3}{National Astronomical Observatory of Japan, Mitaka, Tokyo 181-8588, Japan}
\altaffiltext{4}{Department of Earth and Planetary Science, Kyushu University, Higashi-ku, Fukuoka 812-8581, Japan}

\begin{abstract}
The configuration and evolution of the magnetic field in star-forming cores are investigated in order to directly compare simulations and observations.
We prepare four different initial clouds having different magnetic field strengths and rotation rates, in which magnetic field lines are aligned/misaligned with the rotation axis.
First, we calculate the evolution of such clouds from the prestellar stage until long after protostar formation.
Then, we calculate the polarization of thermal dust emission expected from the simulation data.
We create polarization maps with arbitrary viewing angles and compare them with observations.
Using this procedure, we confirmed that the polarization distribution projected on the celestial plane strongly depends on the viewing angle of the cloud.
Thus, by comparing the observations with the polarization map predicted by the simulations, we can roughly determine the angle between the direction of the global magnetic field and the line of sight.
The configuration of the polarization vectors also depends on the viewing angle.
We find that an hourglass configuration of magnetic field lines is not always realized in a collapsing cloud when the global magnetic field is misaligned with the cloud rotation axis.
Depending on the viewing angle, an $S$-shaped configuration of the magnetic field (or the polarization vectors) appears early in the protostellar accretion phase.
This indicates that not only the magnetic field but also the cloud rotation affects the dynamical evolution of such a cloud.
In addition, by comparing the simulated polarization with actual observations, we can estimate properties of the host cloud such as the evolutionary stage, magnetic field strength, and rotation rate.
\end{abstract}

\section{Introduction}
The magnetic field is believed to play a key role in the star formation process.
Stars form in molecular cloud cores.
In the collapsing cloud core, the magnetic field transfers excess angular momentum and promotes further gas contraction to form a protostar.
After the protostar formation epoch, the protostellar outflow appears to be driven by the magnetic force.
The protostellar outflow influences the interstellar medium and seems to promote subsequent star formation.

To investigate the effects of the magnetic field, we need to determine its strength and morphology.
The strength of the magnetic field in molecular clouds can be determined by Zeeman splitting observations.
Many observations have shown that in a molecular cloud core, the magnetic energy is as large as the gravitational energy \citep[e.g.,][]{crutcher99}.
Thus, the magnetic field in a gravitationally contracting cloud cannot be ignored.
On the other hand, the morphology of the magnetic field in molecular clouds has been obtained by polarization measurements of thermal dust emission.\footnote{Because the dust major axis is aligned perpendicular to the interstellar magnetic field, thermal emission from such dust is linearly polarized, and the B-vector of the electromagnetic wave indicates the direction of the interstellar magnetic field.}
In past studies, the direction of the magnetic field lines at the molecular cloud scale was estimated from the polarization vectors \citep[e.g.,][]{Moneti:1984qy,Tamura:1989uq}.
Recently, \citet{girart06,girart09} obtained a fine structure of collapsing cloud cores by high-angular-resolution measurements of polarized emission. 
They showed a very clear hourglass structure of magnetic field lines (or polarization vectors).
The hourglass structure is believed to be an evidence of gravitational contraction, because the magnetic field, which is well coupled with neutral gas, is pinched toward the center of the gravitationally collapsing gas cloud.

To understand the effects of the magnetic field on the star formation process, information about both the magnetic field strength and its morphology are desirable.
However, it is difficult to determine the magnetic field strength at a small scale (or in a high-density gas region) by resolving the internal structure of the collapsing cloud, which can only be done through Zeeman splitting observations \citep{crutcher2010}.

On the other hand, observation by dust polarization measurement can resolve small-scale structure with (near-future) interferometers such as the Submillimeter Array (SMA) and the Atacama Large Millimeter/submillimeter Array (ALMA).
Thus, we can extract various types of information on the star formation process from polarization observations.

Polarization observations provide information of the magnetic field morphology projected on the celestial plane.
Because, in reality, magnetic field lines have a three-dimensional configuration, we need to reconstruct a three-dimensional configuration of magnetic field lines from the observed polarization vectors, considering the projection effect.
However, such approach is very complicated, and some parameters, such as the projection angle, are uncertain.
Instead, in previous studies, researchers developed models of the magnetic field lines without rotation  \citep{goncalves08, padovani12} or with the effect of rotation \cite[e.g.][]{frau11}.
Then, they compared their model with observations, in which the projection effect of the magnetic field lines are taken into account.
However, since magnetic field lines modeled in such studies are somewhat idealized, it is difficult to investigate effects of different cloud conditions and time evolution on magnetic field lines.
Thus, although such models may roughly explain the large-scale (or molecular cloud scale) magnetic field structure, we need a more sophisticated model based on the cloud collapse simulation to explain recent (or future) high-resolution observations.

Recent observations could resolve the polarization distribution even around the region close to the protostar, where the effect of cloud (or disk) rotation, protostellar outflow, and the circumstellar disk is expected to be significant, in order to investigate (or model) the magnetic field lines.
Thus, we need to construct a more realistic magnetic field line model considering these effects.

The evolution of magnetized clouds has recently been investigated by magnetohydrodynamic (MHD) simulations.
Such simulations reproduced the protostellar outflow \citep{tomisaka02,banerjee06,machida07,hennebelle08a,duffin09} and demonstrated the formation of the circumstellar disk in the collapsing magnetized cloud \citep{hennebelle09, machida11,duffin11}.
These studies also showed the magnetic field lines in three dimensions or the magnetic field vector on an arbitrary cutting plane.
The magnetic field obtained from the numerical simulations is considered to be more realistic than a simple theoretical model.
However, we cannot directly compare the magnetic field obtained from numerical simulations with polarization observations without introducing the effect of projection and thermal dust emission.
Thus, we need a further attempt to compare numerical simulations with observation results.

To compare numerical simulations with observations, \citet{tomisaka11} calculated the polarization of thermal dust radiation from two-dimensional axisymmetric MHD simulation results, assuming a constant emissivity per unit mass.
He called this procedure "observational visualization."
In this study, first, we calculate the cloud evolution from the prestellar cloud core stage until long after protostar formation.
Then, according to the procedure described in \citet{tomisaka11}, we calculate the polarization distribution predicted by MHD simulations and discuss the relationship between the polarization vectors and the evolutionary stage of the cloud or protostar.
\citet{tomisaka11} assumed magnetic field lines parallel to the rotation axis in the initial cloud because he used the data obtained from axisymmetric simulations.
In our study, we investigate both aligned and misaligned magnetic fields.
The initial magnetic field lines are aligned with the initial rotation axis in the former case, whereas they are inclined from the rotation axis in the latter case.
Part of our results was already presented in \citet{shinnaga12}, in which we compared the polarization vectors derived from numerical simulations with polarization observations in a massive star-forming region and determined various cloud parameters and the evolutionary stage of the protostar.
This paper is structured as follows. 
The framework of the MHD simulation is given in \S 2. 
The numerical method for calculating the polarization is presented in \S 3. 
The results of the MHD simulation and polarization calculation are shown in \S4.
We discuss the configuration of magnetic field lines and compare our results with observations in \S 5.
We summarize our results in \S 6.

\section{Model Setting for Cloud Collapse Simulation}
\subsection{Basic Equations and Initial Settings}
\label{sec:model}
To study the evolution of a magnetized cloud core, we solve the three-dimensional resistive MHD equations, including self-gravity:
\begin{eqnarray} 
& \dfrac{\partial \rho}{\partial t}  + \nabla \cdot (\rho \vect{v}) = 0, & \\
& \rho \dfrac{\partial \vect{v}}{\partial t} 
    + \rho(\vect{v} \cdot \nabla)\vect{v} =
    - \nabla P - \dfrac{1}{4 \pi} \vect{B} \times (\nabla \times \vect{B})
    - \rho \nabla \phi, & \\ 	
& \dfrac{\partial \vect{B}}{\partial t} = 
   \nabla \times (\vect{v} \times \vect{B}) + \eta \nabla^2 \vect{B}, & 
\label{eq:reg}\\
& \nabla^2 \phi = 4 \pi G \rho, &
\end{eqnarray}
where $\rho$, $\vect{v}$, $P$, $\vect{B} $, $\eta$, and $\phi$ denote the density, velocity, pressure, magnetic flux density, resistivity, and gravitational potential, respectively.
As the gas pressure, we adopt the isothermal equation of state as
\begin{equation} 
P =  c_{s,0}^2\ \rho,
\label{eq:eos}
\end{equation}
where $c_{s,0} = 190 $\,m\,s$^{-1}$.
The isothermal approximation is justified in the low-density gas region.
\citet{masunaga00} showed that in the collapsing cloud, the gas behaves isothermally for $n\lesssim 10^{10}\cm$ and adiabatically for $n\gtrsim10^{10}\cm$.
In this study, we calculated only a low-density gas region ($n<10^{10} \cm$) using a sink cell treatment (see \S\ref{sec:sink}).
To safely calculate the gas evolution around or inside the sink, we adopted an artificially large resistivity in equation~(\ref{eq:reg}), as
described in \S\ref{sec:sink}.

As the initial state, we assume a spherical cloud with a critical Bonnor--Ebert (BE) density profile $\rho_{\rm BE}$, in which a uniform density is adopted outside the sphere ($r > \rcri$, where $\rcri$ is the critical BE radius).
For the BE density profile, we adopt an isothermal temperature of $T=10$ \,K and a central number density of $n_0 = 5\times10^5 \cm$.
For these parameters, the critical BE sphere has a radius of $\rcri = 6.5\times 10^3$\,AU and a mass of $1.1\msun$.
The gravitational force is ignored outside the host cloud ($r>\rcri$) to mimic a stationary interstellar medium.
Thus, only the gas inside $r<\rcri$ can collapse to form the protostar.
Because the critical BE sphere is in an equilibrium state, we increase the density by a factor of $f=1.67$ to promote contraction, where $f$ is the density enhancement factor.

In the initial cloud, the uniform magnetic field $B_0$ parallel to the $z$-axis is adopted in the entire computational domain.
The cloud rotates rigidly and the rotation axis is inclined with respect to the initial magnetic field lines.
Thus, the initial angular velocity is given by
\begin{eqnarray}
\left(
\begin{array}{l}
\Omega_x \\
\Omega_y \\
\Omega_z \\
\end{array}
\right)
= \Omega_0
\left(
\begin{array}{l}
\mbox{sin}\, \ang \\
0 \\
\mbox{cos}\, \ang \\
\end{array}
\right),
\label{eq:omg-angle}
\end{eqnarray}
where $\ang$ is the angle between the initial magnetic field lines (or $z$-axis) and the rotation axis.

The initial cloud is characterized by three parameters, $B_0$, $\Omega_0$, and $\ang$, which are summarized in the first to third columns of Table~\ref{table:1}.
The non-dimensional quantities $\alpha_0$, $\beta_0$, and $\gamma_0$ are also listed in the forth to sixth columns of Table~\ref{table:1}, where $\alpha_0$, $\beta_0$, and $\gamma_0$ are the ratios of the thermal, rotational, and magnetic energies, respectively, to the gravitational energy inside the initial cloud ($R<R_{\rm c}$).
\begin{table}
\caption{Model parameters}
\label{table:1}
\footnotesize
\begin{center}
\begin{tabular}{c|ccccccccccc} \hline
{\footnotesize Model} & $B_0$ [$\mu$\,G] & $\Omega_0$ {\scriptsize [$\times 10^{-14}\,$s$^{-1}$]} 
 & $\ang$ &   $\alpha_0$ & $\beta_0$ & $\gamma_0$ & $\mu$  \\
\hline
1  & 23  & 0   & 0  &  0.5 & 0    & 0.06 &   6.6  \\
2  & 23  & 14  & 0  &  0.5 & 0.02 & 0.06 &   6.6  \\
3  & 12  & 14  & 60 &  0.5 & 0.02 & 0.02 &  12.6  \\
4  & 72  & 10 & 60 &  0.5 & 0.01 & 0.6 &   2.3  \\
\hline
\end{tabular}
\end{center}
\end{table}

\label{table:1}

We estimate the mass-to-flux ratio of the initial cloud as
\begin{equation}
\dfrac{M}{\Phi} = \frac{M}{\pi R_{\rm cl}^2\, B_0},
\label{eq:mag1}
\end{equation}
where $M$ is the mass contained within the critical radius $R_{c}$ and $\Phi$ is the magnetic flux threading the initial cloud.
A critical value of $M/\Phi$ exists below which a cloud is supported against self-gravity by the magnetic field.
For a cloud with uniform density, \citet{mouschovias76} derived a critical mass-to-flux ratio,
\begin{equation}
\left(\dfrac{M}{\Phi}\right)_{\rm cri} = \dfrac{\zeta}{3\pi}\left(\dfrac{5}{G}\right)^{1/2},
\label{eq:mag2}
\end{equation}
where the constant $\zeta=0.53$ for uniform spheres ($\zeta=0.48$ according to careful calculation by \citealt{tomisaka88a,tomisaka88b}).
For convenience, we use the mass-to-flux ratio normalized by the critical value as
\begin{equation}
\mu \equiv \left(\dfrac{M}{\Phi}\right)/\left(\dfrac{M}{\Phi}\right)_{\rm cri}.
\label{eq:crit}
\end{equation}
The models have the normalized mass-to-flux ratio $\mu$ in the range of $2.3\le\mu\le12.6$.
Thus, the initial clouds are magnetically supercritical.
This is supported by recent observations \citep{crutcher99}.
The normalized mass-to-flux ratio for each model is also listed in the seventh column of Table~\ref{table:1}.

\subsection{Numerical Method and Sink Cells}
\label{sec:sink}
To calculate on a large spatial scale, the nested grid method is adopted \citep[for details, see ][]{machida05a,machida05b}. 
Each level of a rectangular grid has the same number of cells ($ 64 \times 64 \times 64 $).
The calculation is first performed with five grid levels ($l=1-5$).
The box size of the coarsest grid, $l=1$, is set to $2^5\rcri$.
A new finer grid is generated before the Jeans condition is violated \citep{truelove97}.
The box size of the first level of grid ($l=1$) is $L_1 \sim 2.1\times 10^5$\,AU, whereas the maximum grid level, $l_{\rm max} = 9$, has a box size of $L_9=820$\,AU and a cell width of $\Delta_{l=9}= 13$\,AU.

With these settings, we calculated the cloud evolution until a large fraction of the total cloud mass fell onto the protostar. 
To perform long-term calculation of the collapsing cloud, we adopted a sink at the center of the cloud.
We started the calculation without a sink and calculated the cloud evolution for the prestellar gas-collapsing phase without the sink.
Later, we identified protostar formation in the collapsing cloud core when the number density at the cloud center exceeds $n > n_{\rm thr}$, where $n_{\rm thr}$ is the threshold density for our model.
After protostar formation, we calculated the cloud evolution with the sink.

To model the protostar, we adopted a fixed sink with a radius of $r_{\rm sink}=15\,$AU composed of sink cells only around the center of the computational domain.
Because we added no non-axisymmetric perturbation to the initial state, the protostar (or the center of gravity) does not move and remains at the center of the computational domain during the calculation.
In the region $r < r_{\rm sink} = 15\,$AU, gas having a number density of $n > n_{\rm thr} = 3\times10^{9}\cm$ ($\rho_{\rm thr}=1.2\times10^{-14}$\,g\,cm$^{-3}$) is removed from the computational domain and added to the protostar as a gravitating mass in each timestep \citep[for details, see][]{machida09a,machida10}.
Thus, for each timestep, the mass accreted onto the protostar is calculated as
\begin{equation}
\Delta M_{\rm acc} = \int_{r < r_{\rm sink}} [\rho(i,j,k) - \rho_{\rm thr}]\, dV.
\end{equation}

To avoid artificial amplification of the magnetic field around or inside the sink, we adopted the resistivity in equation~(\ref{eq:reg}) as 
\begin{equation}
\eta = c_\eta \dfrac{740}{X_e}\sqrt{\dfrac{T}{10 {\rm K}}} \, \, \, {\rm cm}^2\,{\rm s}^{-1},
\label{eq:etadef}
\end{equation}
where $T$ (10\,K) is the gas temperature and $X_e$ is the ionization degree of the gas, which is expressed as
\begin{equation}
X_e =  5.7 \times 10^{-4} \left( \dfrac{n}{1{\rm cm}^{-3}} \right)^{-1}.
\end{equation} 
Equation~(\ref{eq:etadef}) coincides with the resistivity of the collapsing gas cloud derived in \citet{nakano02} and \citet{machida07} with a numerical factor $c_\eta=1$, in which the magnetic Reynolds number $Re$ becomes $Re<1$ at $n\sim10^{12}\cm$, and the magnetic field begins to dissipate by Ohmic dissipation for $n\gtrsim10^{10}-10^{12}\cm$ \citep[for details, see][]{machida07,machida08a,machida08b}.

The magnetic field is highly amplified around the protostar without the magnetic dissipation because the magnetic flux continues to flow into the sink cell.
Such field may induce artificial effects in the region outside the sink cell.
To avoid this, we chose $c_\eta=10$ to dissipate the magnetic field inside or around the sink; consequently, the magnetic Reynolds number becomes $Re<1$ at $n\sim10^{11}\cm$ and the magnetic field dissipates for $n\gtrsim10^9\cm$.
In our setting, the gas continues to flow into the sink cell, while the magnetic field dissipates around and inside the sink.
We confirmed that this treatment (i.e., artificial dissipation of the magnetic field around the sink) has little effect on the outer gas-collapsing region (see also \S\ref{sec:effect-of-sink}).
Therefore, we can safely calculate the cloud evolution for a long period.

\section{Numerical Method for Polarization Calculation}
\label{sec:pol-calc}
After the MHD simulations were conducted,we calculated the polarization of thermal dust emission using the results of MHD simulations according to the formulae in \citet{tomisaka11} 
 (see \citealt{Fiege00}). 
Here, we briefly summarize the formulation.

The purpose of this study is to calculate the polarization vectors and polarization degrees of thermal dust emission.
For simplicity, we assume that the entire region in the cloud is optically thin and isothermal.
We define the observed direction as the direction of a unit vector $\bf{n}$.
The observational grid normal to $\bf{n}$ has the following horizontal ($\xi$) and vertical ($\eta$) axes:

\begin{equation}
\bf{e}_{\eta}=\left\{
 \begin{array}{ll}
\frac{\bf{e}_{z}-(\bf{e}_{z}\cdot\bf{n})\bf{n}}{|\bf{e}_{z}-(\bf{e}_{z}\cdot\bf{n})\bf{n}|},
 &(\bf{e}_{z}\cdot\bf{n}\ne 0)\\
-\bf{e}_{x}, & (\bf{e}_{z}\cdot\bf{n}=0)
\end{array}\right.
\end{equation}
and
\begin{eqnarray}
\bf{e}_{\xi}=\bf{e}_{\eta}\times\bf{n}.
\end{eqnarray}

Then, we calculated the relative Stokes parameters,
\begin{equation}
q=\int\rho\,  \cos2\psi\, \cos^{2}\gamma \, ds
\label{eq:stokesq}
\end{equation}
and
\begin{equation}
u=\int\rho\,  \sin2\psi\, \cos^{2}\gamma \, ds,
\end{equation}
where the integration is performed along the line of sight $\bf{n}$, and the angles $\psi$ and $\gamma$ are between the projected magnetic field direction and the $\eta$-axis and between the magnetic field direction and the $\xi \eta$ plane, respectively.
The polarization degree is given as
\begin{eqnarray}
P=p_{0}\frac{(q^{2}+u^{2})^\frac{1}{2}}{\Sigma-p_{0}\Sigma_{2}}
\label{eq:pol}
\end{eqnarray}
from two quantities obtained by integration of the density along the line of sight as
\begin{eqnarray}
\Sigma=\int\rho\, ds
\end{eqnarray}
and
\begin{eqnarray}
\Sigma_{2}=\int\rho \left( \frac{\cos^{2}\gamma}{2}-\frac{1}{3} \right) ds.
\label{eq;sigma2}
\end{eqnarray}
In equation~(\ref{eq:pol}), the numerical factor $p_{0}$ is set to $0.15$ in order to fit the observational maximum polarization degree in interstellar space.

\begin{figure}
\begin{center}
\includegraphics[width= 80mm]{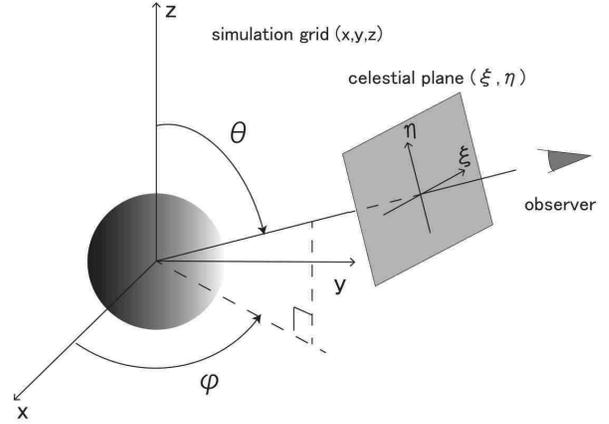}
\end{center}
\caption{
Relationship between simulation grid and observation grid.
}
\label{fig:grid}
\end{figure}

Using equations~(\ref{eq:stokesq})-(\ref{eq;sigma2}), we calculated the polarization at various viewing angles for different datasets taken from simulations.
The directions of the viewing angles are specified by the angles $\theta$ and $\phi$: $\theta$ represents the inclination angle and $\phi$ is the azimuthal angle.
The case of $\theta = 0\degr$ corresponds to the pole-on view, whereas $\theta = 90\degr$ corresponds to the edge-on view.
Because of symmetry, we examine the inclination angle in the range of $0\degr \le \theta \le 90\degr$.
The relationship between the simulation grid ($xyz$ coordinate system) and the observation grid ($\xi\eta$ coordinate system) is shown in Figure~\ref{fig:grid}. 

\section{Results}

In this section, we show the results of MHD simulations in \S\ref{sec:simulation} and their observational visualizations in \S\ref{sec:visualization} for the four models (models 1-4) listed in Table~\ref{table:1}.
We use different coordinate systems in \S\ref{sec:simulation} and \S\ref{sec:visualization} because we observe the collapsing cloud from an arbitrary direction in \S\ref{sec:visualization}.

\begin{figure}
\begin{center}
\includegraphics[width= 80mm]{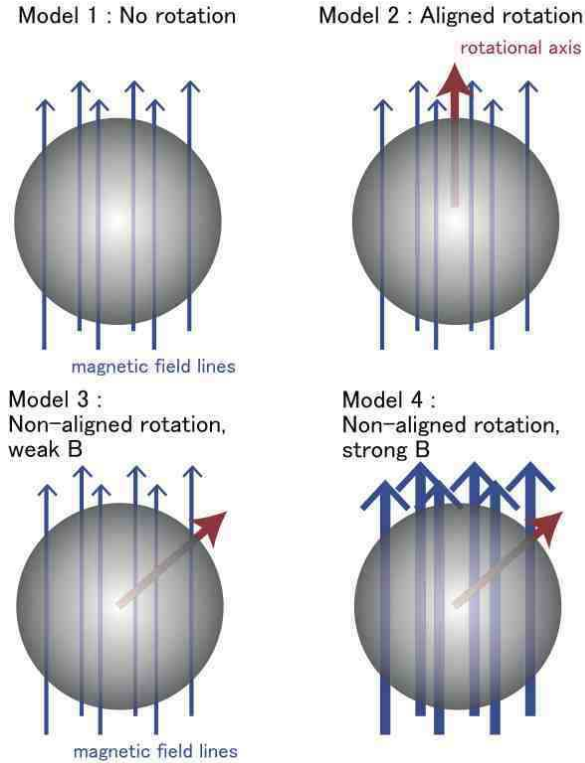}
\end{center}
\caption{
Schematic view of the initial conditions for each model. 
Blue and red arrows represent the directions of initial magnetic field lines and initial rotation axis, respectively.
The width of the blue arrows reflects the magnetic field strength.
The magnetic field lines are aligned with the rotation axis for model 2, whereas they are inclined from the rotation axis by $\delta_0 =60\degr$ for models 3 and 4.}
\label{fig:initial}
\end{figure}

The configurations and directions of the magnetic field lines and rotation axes for the models are schematically illustrated in Figure~\ref{fig:initial}.
For all the models, we adopted a uniform magnetic field that has the same direction but different strengths for each model.
As seen in the figure, model 1, which is the standard model, has a uniform magnetic field parallel to the $z$-axis without rotation.
Model 2 has a magnetic field of the same strength and configuration but with rotation; the rotation axis is parallel to the magnetic field lines (or parallel to the $z$-axis).
On the other hand, the rotation axis is inclined from the magnetic field lines at an angle of $\ang=60\degr$ in models 3 and 4.
Model 3 has a weaker magnetic field than model 2, whereas model 4 has the strongest magnetic field among all the models, as shown in Table~\ref{table:1}.
In addition, the rotation rate of model 4 is slightly slower than those of models 2 and 3.

\subsection{Magnetohydrodynamic Simulations}
\label{sec:simulation}
\begin{figure}
\begin{center}
\includegraphics[width= 80mm]{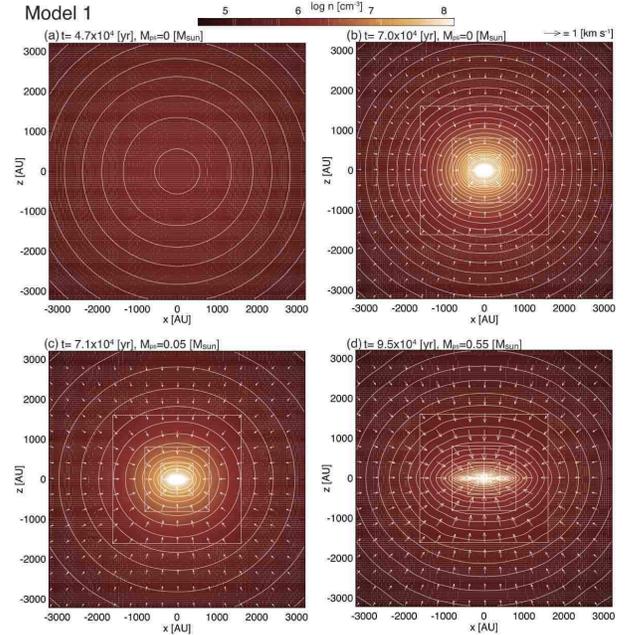}
\end{center}
\caption{
Density distribution (colors and contours) and velocity vectors (arrows) on the $y=0$ cutting plane at four different epochs for model 1.
Elapsed time $t$ and protostellar mass $M_{\rm ps}$ are listed in each panel.
}
\label{fig:a0b0}
\end{figure}

In this subsection, we simply show the results of MHD simulations.
As described in \S\ref{sec:model}, we calculated the evolution of the magnetized collapsing cloud from the prestellar cloud stage.
Figure~\ref{fig:a0b0} shows the cloud evolution for model 1, in which the density distribution (colors and contours) and velocity vectors on the $y=0$ cutting plane of the $l=7$ grid are plotted.
Figure~\ref{fig:a0b0}(a) shows the initial state.
Figure~\ref{fig:a0b0}(b) shows the collapsing cloud before protostar formation (i.e., the prestellar stage), and Figures~\ref{fig:a0b0}(c) and (d) show the cloud after protostar formation (i.e., the protostellar stage).
In this model, a protostar forms at $t=7.1\times 10^4$\,yr after the cloud begins to collapse.
At the epoch corresponding to Figure~\ref{fig:a0b0}(d), the protostar has a mass of $0.55\msun$ and is still embedded in the collapsing cloud.
Because the initial cloud in model 1 has a magnetic field but no rotation, the gas collapses essentially along the magnetic field lines.
Thus, a disk-like structure (pseudo-disk) forms, extending in the direction perpendicular to the magnetic field lines, as seen in Figure~\ref{fig:a0b0}(d).

\begin{figure}
\begin{center}
\includegraphics[width= 80mm]{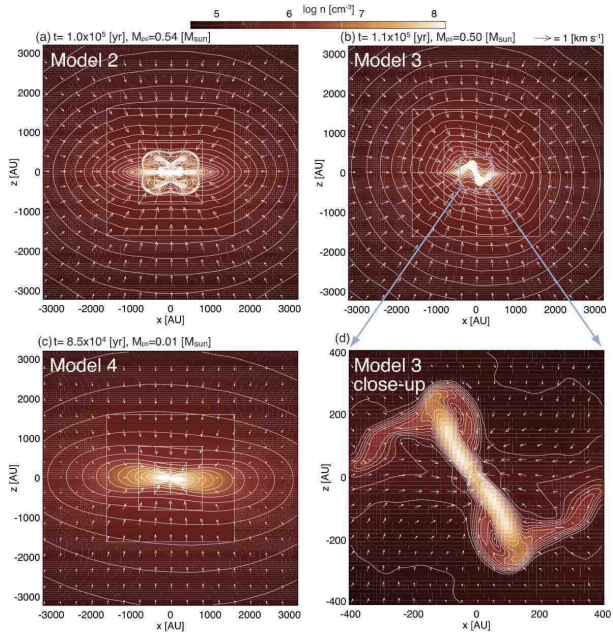}
\end{center}
\caption{
Density distribution (colors and contours) and velocity vectors (arrows) on the $y=0$ cutting plane for models 2 (a), 3 (b and c), and 4 (c).
Elapsed time $t$ and protostellar mass $M_{\rm ps}$ are listed in each panel.
Panel (d) is a close-up view of panel (b).
}
\label{fig:allmodels}
\end{figure}

Figure~\ref{fig:allmodels} shows snapshots of the protostellar stage for models 2, 3, and 4.
After protostar formation, model 2 shows a weak outflow that is traced as a shock front near $z\sim 800$\,AU, far from the center of the cloud in Figure~\ref{fig:allmodels}(a).
Note that we cannot resolve the outflow driven by the circumstellar disk with sufficient spatial resolution because we adopted sink cells in the high-density gas region in order to perform a long-term calculation (see \S\ref{sec:effect-of-sink}).
The large- and small-scale structures for model 3 are plotted in Figures~\ref{fig:allmodels}(b) and (d).
For this model, the disk is not perpendicular to the $z$-axis, that is, the disk normal is inclined from the direction of the global magnetic field lines.
This is because the rotation axis of the initial cloud is inclined from the initial magnetic field lines.
In addition, for model 3, the initial rotational energy ($\beta_0=0.02$) is equivalent to the initial magnetic energy ($\gamma_0=0.02$). 
Thus, the cloud's rotation can affect the dynamical evolution of the cloud and the disk is inclined from the $z$-axis, as seen in Figure~\ref{fig:allmodels}(d).	
On the other hand, although the rotation axis is inclined from the magnetic field direction in the initial cloud, the disk tends to form in the direction perpendicular to the magnetic field lines for model 4 (Figure~\ref{fig:allmodels}[c]).
For this model, because the initial magnetic energy ($\gamma_0=0.57$) dominates the initial rotational energy ($\beta_0=0.01$), the cloud rotation is expected to have very little effect on the dynamical cloud evolution \citep[for details, see][]{machida06}.
In addition, because of the strong magnetic field, magnetic braking is more effective in model 4 than in models 2 and 3.

For all the models, over half of the initial cloud mass falls onto either the protostar or the circumstellar disk by the end of the calculation.
This means that we calculated the cloud (or protostar) evolution until the Class 0 or I protostar stage.
We will show the evolution of the magnetic field configuration in the next subsection.

\subsection{Expected Distributions of Polarization}
\label{sec:visualization}
After MHD simulations were performed, we calculated the polarization of the thermal dust emission according to the procedure described in \S\ref{sec:pol-calc} and \citet{tomisaka11}.
In this subsection, we show the expected distributions of the polarization.
Some of the arguments in the next subsection are similar to previous studies \citep{frau11, goncalves08}.
However, to compare models with and without rotation, or compare aligned rotation model with misaligned rotation model, we describe polarization patterns for all models in detail.

\subsubsection{Model 1: Non-rotating Cloud}
\label{sec:standard}
\begin{figure}
\begin{center}
\includegraphics[width= 80mm]{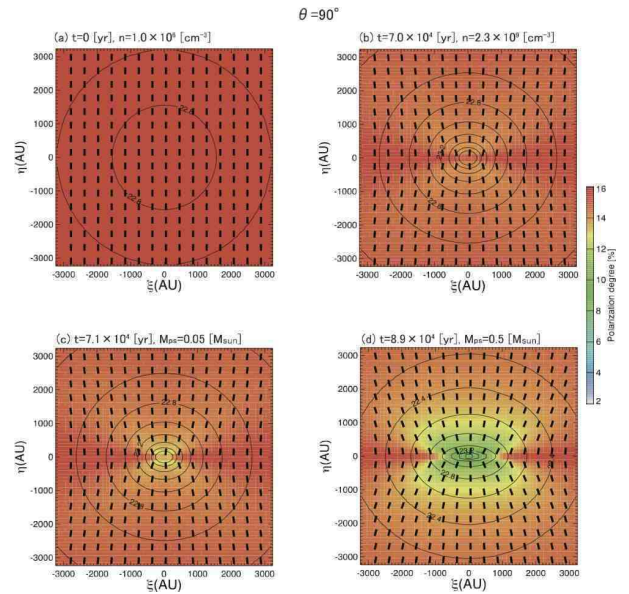}
\end{center}
\caption{
Polarization maps for model 1 viewed from the angle ($\theta$, $\phi$) = ($90\degr$, $0\degr$)
 at four different epochs, which correspond to the respective panels of Figure~\ref{fig:a0b0}.
The polarization degree (colors), polarization vector (lines), and column density (contours) are plotted in each panel.
The number density at the center of the cloud $n_{\rm c}$ or protostellar mass $M_{\rm ps}$ is shown in each panel.
}
\label{fig:standard_time}
\end{figure}

Figure~\ref{fig:standard_time} plots the polarization distribution for model 1 at different epochs viewed from the angle of $(\theta, \phi) = (90\degr, 0\degr)$.
In this model, the initial cloud is assumed to have a uniform magnetic field but no rotation.
In the figure, the colors, black contours, and straight lines represent the polarization degrees, column densities, and polarization vectors (the direction of the B-vector of the electromagnetic waves), respectively.
Hereafter, we call this type of figure a polarization map.
Viewed from $(\theta, \phi) = (90\degr, 0\degr)$, the $\xi$ and $\eta$ directions in the observation grid correspond to the $y$ and $z$ directions in the simulation grid, respectively (see Figure~\ref{fig:grid}).
Thus, each panel of Figure~\ref{fig:standard_time} corresponds to the same panel in Figure~\ref{fig:a0b0}.
(Note that the distributions on the $y=0$ and $x=0$ planes are the same because the system is axisymmetric.
In addition, because this model maintains the axisymmetry, the physical quantities do not depend on the azimuthal viewing angle $\phi$.
Thus, we show only the polarization map with $\phi=0\degr$ for this model.)

Figure~\ref{fig:standard_time}(a) is a polarization map of the initial state for model 1, in which all the polarization vectors are aligned with the $\eta$-axis.
Figure~\ref{fig:standard_time}(b) shows the polarization just before protostar formation.
At this epoch, the polarization vectors are distorted near the center of the cloud, whereas the column density contours maintain a spherical symmetry.
Figure~\ref{fig:standard_time}(c) shows the polarization after protostar formation.
The polarization degree around the protostar (or around the center of the cloud) gradually decreases because the magnetic field parallel to the line of sight begins to contribute as the cloud collapses. 
After protostar formation, the magnetic field lines point to the protostar around the center of the cloud, generating a magnetic field parallel to the line of sight.
As the parallel component increases, the polarization degree decreases, because only the magnetic field lines perpendicular to the line of sight contribute to the polarization.
(We assume that the magnetic field changes its direction while maintaining its strength.) 
Around the center of the cloud, the polarization degree decreases to $\sim13$\% when the protostellar mass is $M_{\rm ps} =0.05\msun$ (Figure~\ref{fig:standard_time}[c]), and to $<10$\% when $M_{\rm ps} = 0.5\msun$ (Figure~\ref{fig:standard_time}[d]).
This indicates that the magnetic field lines are gradually inclined toward the line-of-sight direction. 

From the polarization vectors in the $\xi\eta$ plane, we can directly confirm that the magnetic field lines are inclined from their initial direction.
In other words, the magnetic field lines point to the protostar around the center of the cloud.
Thus, we can imagine an hourglass structure of the magnetic field lines on the polarization maps.
In addition, the column density contours show a disk-like configuration, which indicates the existence of a pseudo-disk, as described in \S\ref{sec:simulation}.

\begin{figure*}
\begin{center}
\includegraphics[width=120mm]{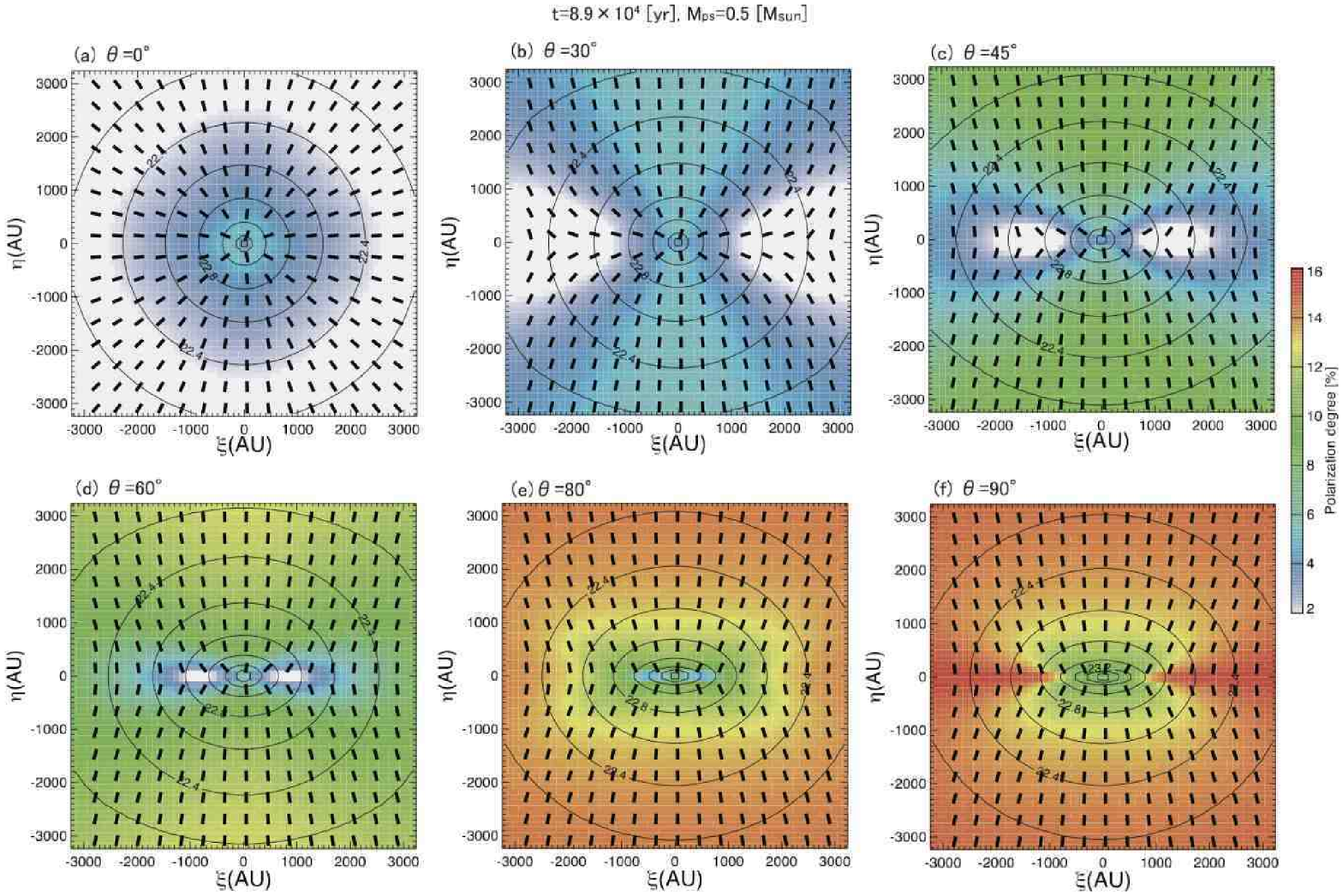}
\end{center}
\caption{
Polarization maps for model 1 viewed from $\theta =0\degr$ (pole-on view; a), $30\degr$ (b),
 $45\degr$ (c), $60\degr$ (d), $80\degr$ (e), and $90\degr$ (edge-on view; f); $\phi=0\degr$ when $M_{\rm ps}=0.5\msun$.}
\label{fig:standard_theta}
\end{figure*}

To investigate the effects of viewing angle $\theta$ on the polarization, we selected a snapshot when the protostar has a mass of $M_{\rm ps} = 0.5\msun$. 
Figure~\ref{fig:standard_theta} shows polarization maps with different viewing angles of $\theta=0\degr$ (a), $30\degr$ (b), $45\degr$ (c), $60\degr$ (d), $80\degr$ (e), and $90\degr$ (f); the azimuthal angle is fixed at $\phi=0\degr$. 
The viewing angle $\theta=0\degr$ corresponds to the pole-on view and $\theta=90\degr$ corresponds to the edge-on view.
Thus, Figure~\ref{fig:standard_theta}(f) is identical to Figure~\ref{fig:standard_time}(d).
Figure~\ref{fig:standard_theta} indicates that the properties of the polarization map depend strongly on the viewing angle $\theta$.
The figure shows that the polarization degree of the $\theta=0\degr$ map (panel [a]) is considerably lower than those for $\theta \ne 0\degr$. 
We assumed that the initial magnetic field lines are parallel to the $z$-axis.
Thus, as the cloud evolves, the $z$ component of the magnetic field ($B_z$) dominates the other components ($B_r$ and $B_\phi$) in the entire cloud.
Therefore, the lower polarization is a natural consequence of the configuration when we measure the polarization from the pole (or we integrate the polarization along the $z$-axis) because the magnetic field lines are almost parallel to the line of sight.
In Figure~\ref{fig:standard_theta}(a), the central region has a larger polarization degree, which indicates that magnetic field lines perpendicular to the line of sight exist around the center of the collapsing cloud.
In MHD simulations, the azimuthal component of the magnetic field ($B_\phi$) is not generated because model 1 has no rotation, whereas the radial component ($B_r$) is generated as the cloud collapses.
Because the magnetic field lines are strongly distorted near the protostar, the central region has a larger polarization degree owing to the non-negligible component of $B_r$ in Figure~\ref{fig:standard_theta}(a).

\begin{figure}
\begin{center}
\includegraphics[width=70 mm]{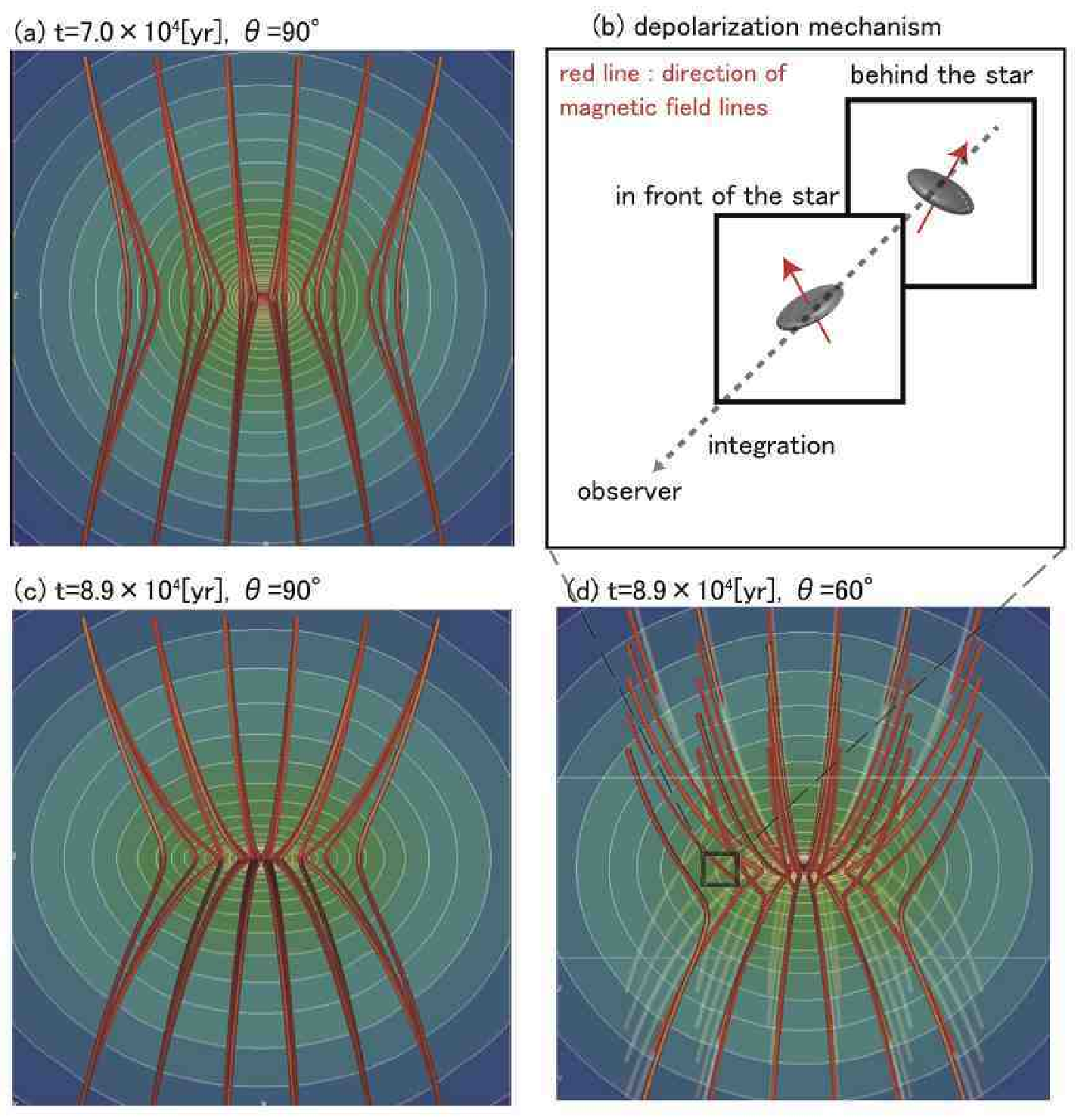}
\end{center}
\caption{
Magnetic field lines in three dimensions before (a) and after (c) protostar formation for model 1.
In panel (d), $\theta=60\degr$, and the two panels on the left show the edge-on view ($\theta = 0\degr$).
Red lines represent magnetic field lines traced from the lower and upper boundary surfaces. 
White contour lines and colors represent the density distribution on the $y=0$ cutting plane.
Panel (b) gives a schematic explanation of the depolarization mechanism
 when we observe the cloud from $\theta=60\degr$.
In the region marked in panel (d),
 the foreground and background magnetic field lines intersect at a large angle $\simeq 90\degr$.
In this case, dust alignment between the foreground and background is canceled,
 which decreases the polarization degree. 
}
\label{fig:standard_AVS}
\end{figure}

On the polarization maps for $\theta =30\degr$, $45\degr$, and $60\degr$ in Figures~\ref{fig:standard_theta}(b), (c), and (d), respectively, an interesting feature appears around the protostar.
In these maps, a region of very low polarization with a polarization degree of $\lesssim 2$\% (white region) appears in the transverse direction around the protostar (or along the $\xi$-axis).
The low-polarization area shrinks as the viewing angle $\theta$ increases.
Finally, the region disappears on the polarization maps of $\theta=80\degr$ (Figure~\ref{fig:standard_theta}[e]) and $90\degr$ (Figure~\ref{fig:standard_theta}[f]).
Thus, this low polarization is attributed to the projection effect of the magnetic field lines. 
To compare the projected magnetic field lines on an arbitrary plane with those in three dimensions, a three-dimensional view of the magnetic field lines is plotted in Figure~\ref{fig:standard_AVS}.
Figure~\ref{fig:standard_AVS}(a) shows the magnetic field lines before protostar formation (corresponding to Figure~\ref{fig:standard_time}[b]), whereas panels (c) and (d) show those after protostar formation (corresponding to Figure~\ref{fig:standard_time}[d]).
In addition, panel (d) has a viewing angle of $\theta=60\degr$, whereas the left panels, (a) and (c), have $\theta=90\degr$ (i.e., an edge-on view).

The polarization degree is lower around the protostar because of two reasons.
First, the magnetic field parallel to the line of sight does not contribute to the integration of the polarization.
The magnetic field lines are strongly bent toward the protostar around the center of the cloud.
As seen in Figure~\ref{fig:standard_AVS}, these field lines have a component parallel to the line of sight.
Thus, the polarization degree decreases near the center of the cloud.
The other reason is the cancelation of positive Stokes $(q,u)$ and negative Stokes $(q,u)$ along one line of sight (see Figure~\ref{fig:standard_AVS}[c]).
That is, two overlapping magnetic field lines cancel the polarization if they are perpendicular to each other because the direction of dust alignment is perpendicular.
The model has an axisymmetric magnetic field with respect to the $z$-axis and a line-symmetric distribution with respect to the $z=0$ plane.
In this case, as shown in Figure~\ref{fig:standard_AVS}(d), the foreground magnetic field (red) and the background field (blue) overlap each other with an angle of $\lesssim 90\degr$.
This weakens the polarization in the polarization map.
This occurs in the region $z\simeq 0$ or $\eta\sim 0$ in a symmetric way with respect to the $z$-axis or $\eta$-axis in this case.

In the collapsing cloud, the magnetic field lines have an hourglass-shaped configuration.
As shown in Figures~\ref{fig:standard_AVS}(a) and (c), a clear hourglass appears when we see the magnetic field lines from the edge (the edge-on view, $\theta=90\degr$). 
On the other hand, a complicated structure is expected around the center of the cloud when viewed from $\theta < 90\degr$ (Figure~\ref{fig:standard_AVS}[d]).
In reality, we cannot observe the three-dimensional structure of the magnetic field lines; we observe only the projected magnetic field onto the polarization map.
Instead, Figure~\ref{fig:standard_theta} indicates that we can roughly estimate the angle between the global magnetic field lines and the line of sight from the distribution of the low-polarization region.

\begin{figure}
\begin{center}
\includegraphics[width= 80mm]{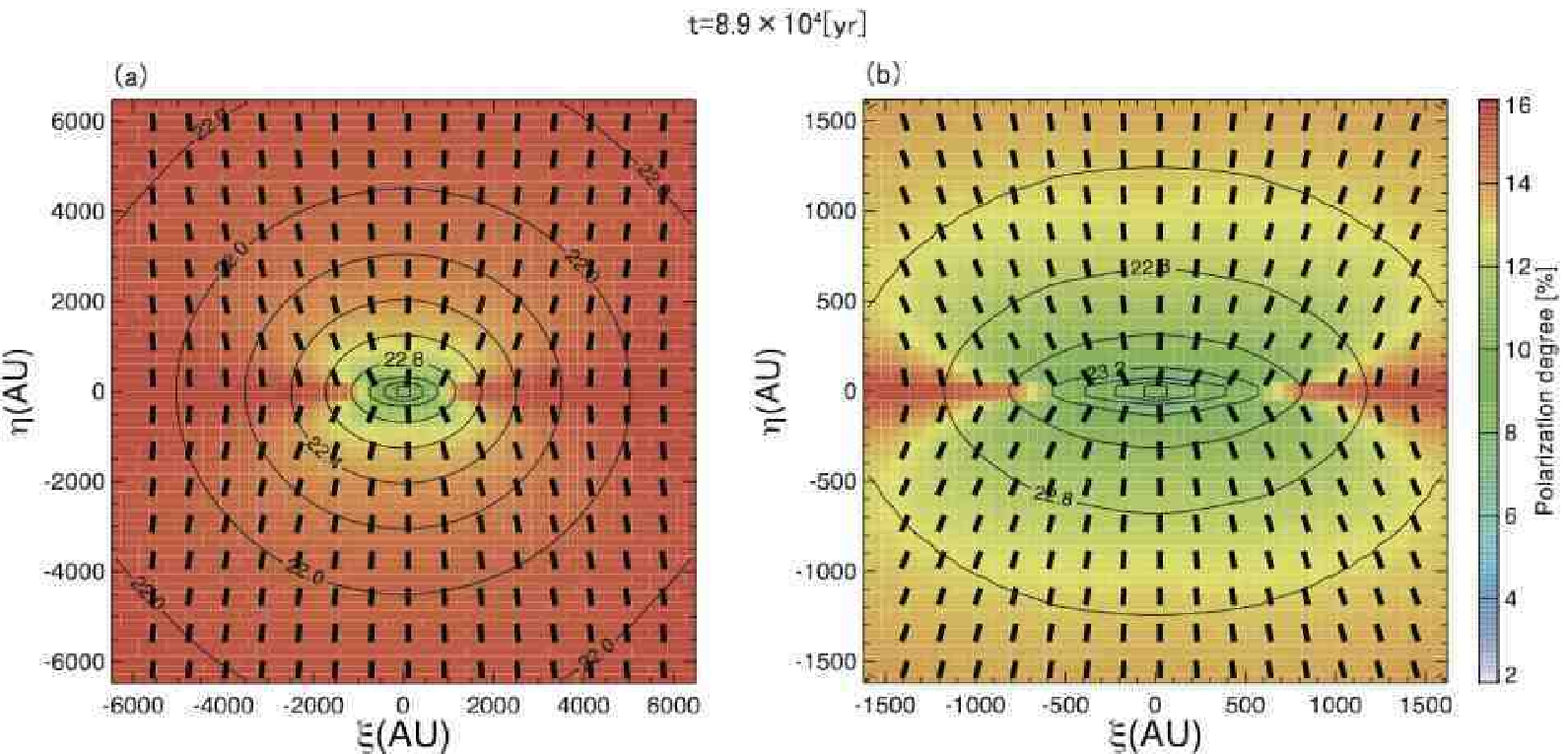}
\end{center}

\caption{
Polarization maps for model 1 at different spatial scales with ($\theta$, $\phi$) = ($90\degr$, $0\degr$) when $M_{\rm ps}=0.5\msun$.
Right panel shows the central region in the left panel enlarged four times.
}
\label{fig:standard_scale}
\end{figure}

We also investigated the scale dependency of the configuration of the magnetic field lines.
Figure~\ref{fig:standard_scale} shows polarization maps with $\theta=90\degr$ at different spatial scales.
Panel (a) is 12000\,AU$\times$12000\,AU in size, whereas panel (b) is 3000\,AU$\times$3000\,AU in size.
Figure~\ref{fig:standard_scale} clearly indicates that the polarization vectors closer to the protostar are more distorted.
Thus, the hourglass structure is emphasized at smaller scales.

\subsubsection{Model 2: Rotating Cloud}
\label{sec:rotation}
\begin{figure}
\begin{center}
\includegraphics[width=80mm]{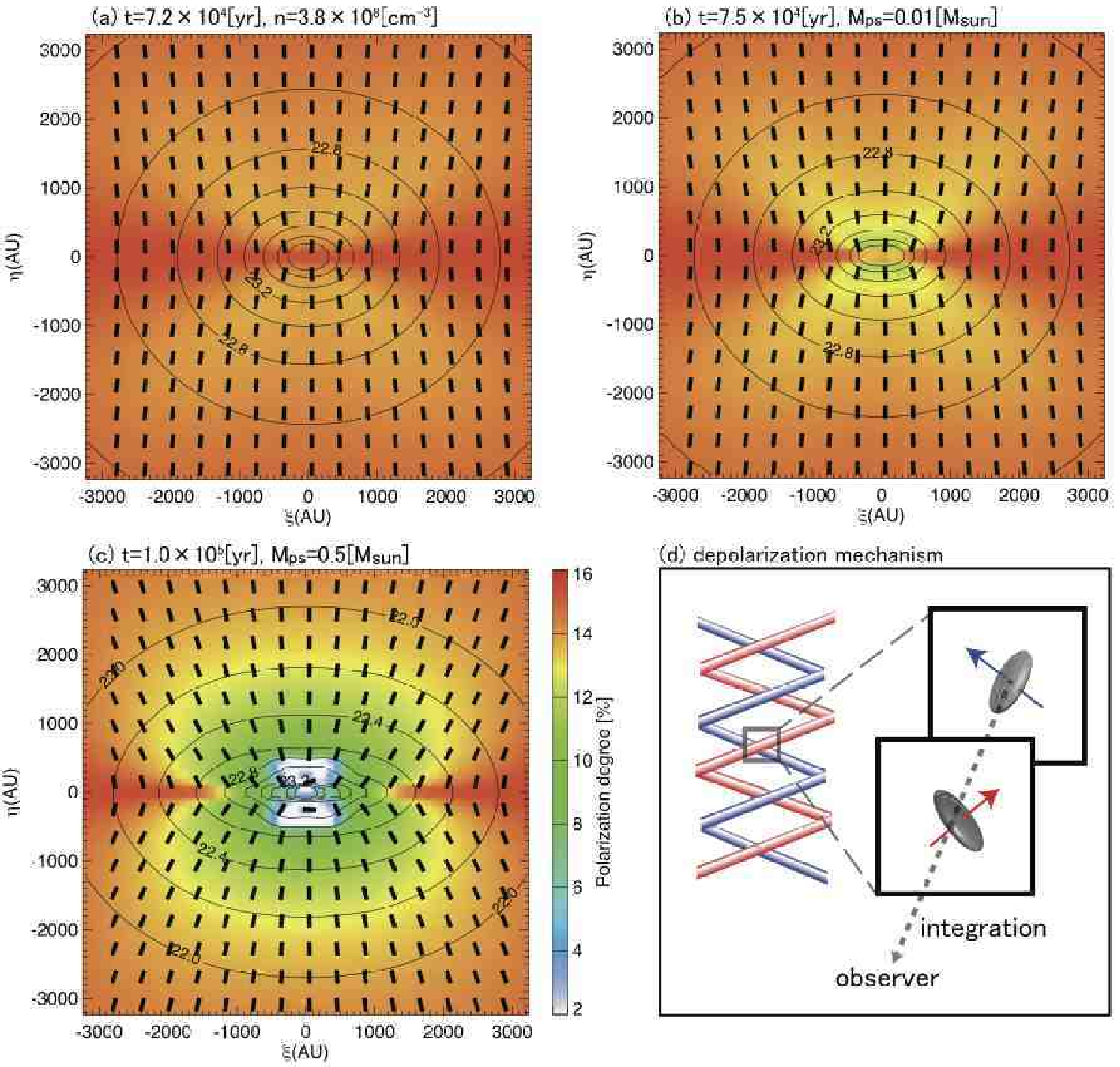}
\end{center}
\caption{
Polarization maps for model 2 at three different epochs 
 ($t=7.2\times 10^4 {\rm yr}$, $7.5\times 10^4 {\rm yr}$, and $1\times 10^5 {\rm yr}$)
 at the same viewing angle of ($\theta$, $\phi$) = ($90\degr$, 0$\degr$).
Panel (a) represents the prestellar stage, and
 panels (b) and (c) show the protostellar stage.
The protostar masses at each snapshot are
 $M_{\rm ps}=0.01\msun$ (panel [b]) and $M_{\rm ps}=0.5\msun$ (panel [c]).
}
\label{fig:rotation_time}
\end{figure}

To investigate the effect of cloud rotation on the polarization, we calculated the polarization for model 2, which has the same magnetic field strength as model 1 but a finite angular momentum.
In model 2, the rotation axis is parallel to the magnetic field lines, as shown in Figure~\ref{fig:initial}.
Figure~\ref{fig:rotation_time} shows the polarization maps for model 2 at different epochs for $(\theta,\phi)=(90\degr,0\degr)$. 
Panels (a) and (b) show the polarization before and just after protostar formation, respectively.
At these epochs, the polarization distribution for model 2 is almost the same as that for model 1 (compare Figure~\ref{fig:rotation_time} with Figure~\ref{fig:standard_time}).
All the panels of this figure indicate an hourglass configuration of the magnetic field lines.
In addition, a region of slightly low polarization (orange region) appears above and below the center of the cloud, as seen in Figure~\ref{fig:standard_time}.
Thus, the cloud rotation has very little effect on the polarization distribution for the edge-on view before and just after protostar formation.

On the other hand, after protostar formation, the polarization degree of model 2 is very different from that of model 1 with no rotation. 
As shown in Figure~\ref{fig:rotation_time}(c), in the rotating cloud, the polarization degree just above and below the protostar is extremely low ($<2$\%, white region). 
Even in model 1 (non-rotating cloud), a low-polarization region appears around the protostar (Figures~\ref{fig:standard_time}[c] and [d]).
However, the polarization degree around the protostar is $\sim10\%$ in model 1 but $<2\%$ in model 2.
In the non-rotating model (model 1), the low polarization degree is caused by the generation of the radial component of the magnetic field, which produces magnetic field lines parallel to the line of sight when viewed edge-on. 
On the other hand, in the rotating cloud, in addition to the radial component, a toroidal component ($B_\phi$) is generated by the rotation.
Because the system is axisymmetric with respect to the $z$-axis, the angle between the foreground and background magnetic field lines is large ($\sim90\degr$) when the magnetic field consists of poloidal and toroidal components of comparable strength (see Figure~\ref{fig:rotation_time}[d]). 
Consequently, a region of extremely low polarization appears above and below the protostar, where the toroidal component is amplified to become comparable with the poloidal one.
This region appears only after protostar formation because the toroidal magnetic field is a result of rotational motion around the protostar.
Note that in the isothermal collapse phase (i.e., before protostar formation), such a strong toroidal field is not expected because the collapse timescale is shorter than the rotational timescale.
Therefore, the extremely low polarization degree around the center of the cloud is an indicator of the existence of the protostar and rotating disk.

\begin{figure*}
\begin{center}
\includegraphics[width= 120mm]{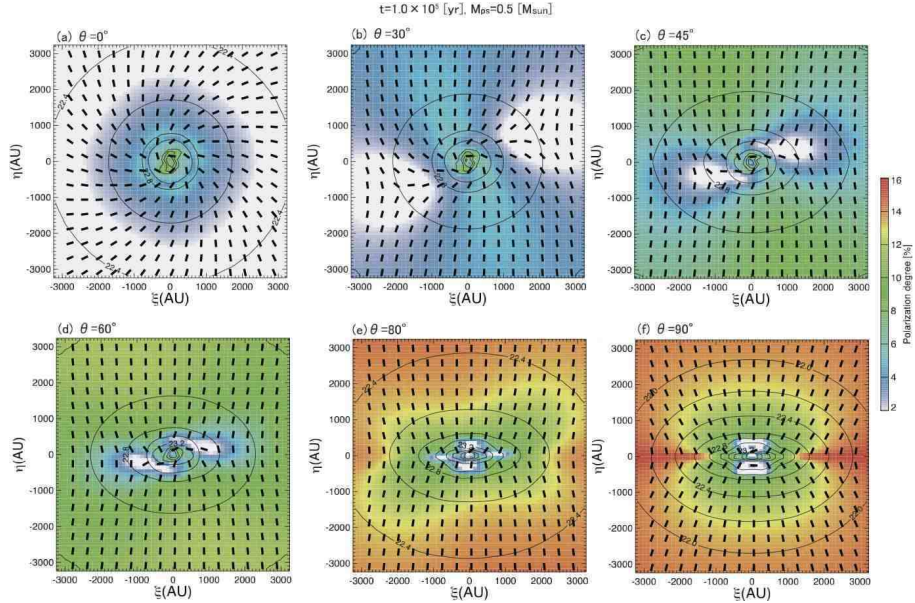}
\end{center}
\caption{
Polarization maps for model 2 from the viewing angles of $\theta =0\degr$ (pole-on view; a),
 $30\degr$ (b), $45\degr$ (c), $60\degr$(d), $80\degr$ (e), and $90\degr$ (edge-on view; f);
 $\phi=0\degr$ when $M_{\rm ps}=0.5\msun$.
}
\label{fig:rotation_theta}
\end{figure*}

Figure~\ref{fig:rotation_theta} shows polarization maps with different viewing angles ($\theta = 0\degr$ [a], $30\degr$ [b], $45\degr$ [c], $60\degr$ [d], $80\degr$ [e], and $90\degr$ [f]; $\phi=0\degr$) for model 2 when the protostar has a mass of $M_{\rm ps} = 0.5\msun$.
Similar to model 1, because model 2 maintains axisymmetry with respect to the $z$-axis, the polarization distribution does not depend on the azimuthal viewing angle $\phi$.
A comparison of Figures~\ref{fig:rotation_theta} (rotating model) and \ref{fig:standard_theta} (non-rotating model) reveals a difference in the polarization distribution.
In the non-rotating model, a low-polarization region ($<2$\%) appears in a transverse direction ($\xi$-direction) around the protostar at viewing angles of $\theta = 30\degr$, $45\degr$, and $60\degr$ (Figure~\ref{fig:standard_theta}).
This feature was explained using Figure~\ref{fig:standard_AVS} in \S\ref{sec:standard}.
On the other hand, in the rotating model, the low polarization degree appears in two regions, upper-right and lower-left regions of the center, as shown in the polarization maps with $\theta = 30\degr$, $45\degr$, and $60\degr$ in Figure~\ref{fig:rotation_theta}.
Thus, in model 2, the line symmetry of the polarization distribution with respect to the $\eta$-axis is broken and a point symmetry with respect to the protostar appears. 
This point-symmetric distribution of the polarization is a striking feature of rotating clouds \citep{tomisaka11}.
\citet{tomisaka11} pointed out that (i) in the protostellar phase, the toroidal magnetic field arises owing to rotation of the central disk.
This toroidal component has antisymmetry with respect to the $z=0$ plane (for example, $B_\phi>0$ for $z>0$, and $B_\phi<0$ for $z<0$).
(ii) The magnetic field consisting of the toroidal and hourglass-type poloidal components becomes point-symmetric (see his Figure 10).
When a magnetic field of this type is seen from $30\degr\lesssim \theta \lesssim 60\degr$, the region with a low polarization degree is distributed in a point-symmetric way.
In other words, this is caused by a combination of viewing angle and rotation effects.

As described above, a strong toroidal field also causes a low polarization degree.
This low polarization appears in the region just above and below the protostar and exhibits line symmetry with respect to the $\xi$-axis, as shown in Figure~\ref{fig:rotation_theta}(f).
Thus, a low-polarization region just above and below the central protostar appears in the $\theta=45\degr$, $60\degr$, $80\degr$, and $90\degr$ maps in Figures~\ref{fig:rotation_theta}(c), (d), (e), and (f), respectively, and is caused by the coexisting toroidal and poloidal magnetic field components.
In summary, two effects decrease the polarization degree: line-symmetric and point-symmetric effects.
Both arise from cancelation between the foreground and background dust alignments.
Even without the toroidal field (model 1), cancelation in the poloidal component produces a low-polarization region extending near the $\xi$-axis.
On the other hand, a magnetic field consisting of poloidal and toroidal components causes vertical extension of the low-polarization region near the $\eta$-axis.
In this case, the former cancelation occurs in the upper-right and lower-left regions of the center in a point-symmetric way when viewed at angles in the range of $30\degr \lesssim \theta \lesssim 60\degr$.

\begin{figure}
\begin{center}
\includegraphics[width= 80mm]{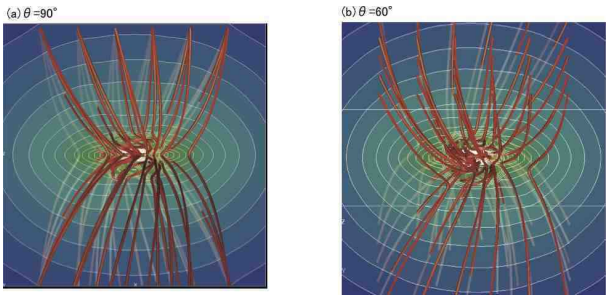}
\end{center}
\caption{
Magnetic field lines in three dimensions when $M_{\rm ps} = 0.5\msun$ for model 2.
Red lines represent magnetic field lines.
White contours and colors represent density distribution on the $y=0$ cutting plane.
The viewing angles are $\theta=90\degr$ (a) and $\theta=60\degr$ (b).
}

\label{fig:compare_rotation}
\end{figure}

To clarify the effects described above, we plot a three-dimensional view of the magnetic field lines in Figures~\ref{fig:compare_rotation}(a) and (b) at viewing angles of $\theta=90\degr$ and $\theta = 60\degr$, respectively.
The figure clearly shows that the toroidal field is confined in a small region around the protostar, and the hourglass configuration of the magnetic field lines extends to the large scale.

Although the polarization distribution in the rotating model differs from that in the non-rotating model (compare Figure~\ref{fig:standard_theta} with Figure~\ref{fig:rotation_theta}), the polarization vectors and column density contours of the models are almost the same at the large scale.
In addition, the hourglass configuration at the cloud scale can be confirmed from the polarization maps in both models.
At the small scale, however, the polarization vectors of the rotating model deviate from the hourglass configuration near the protostar because the toroidal component dominates the poloidal one.
In terms of observations, it is difficult to observe the deviation because the polarization degree is too low to allow identification of the direction of the polarization in such a region.

\subsubsection{Model 3: Misaligned Cloud with Weak Magnetic Field}
\label{sec:weakB}
\begin{figure*}
\begin{center}
\includegraphics[width= 120mm]{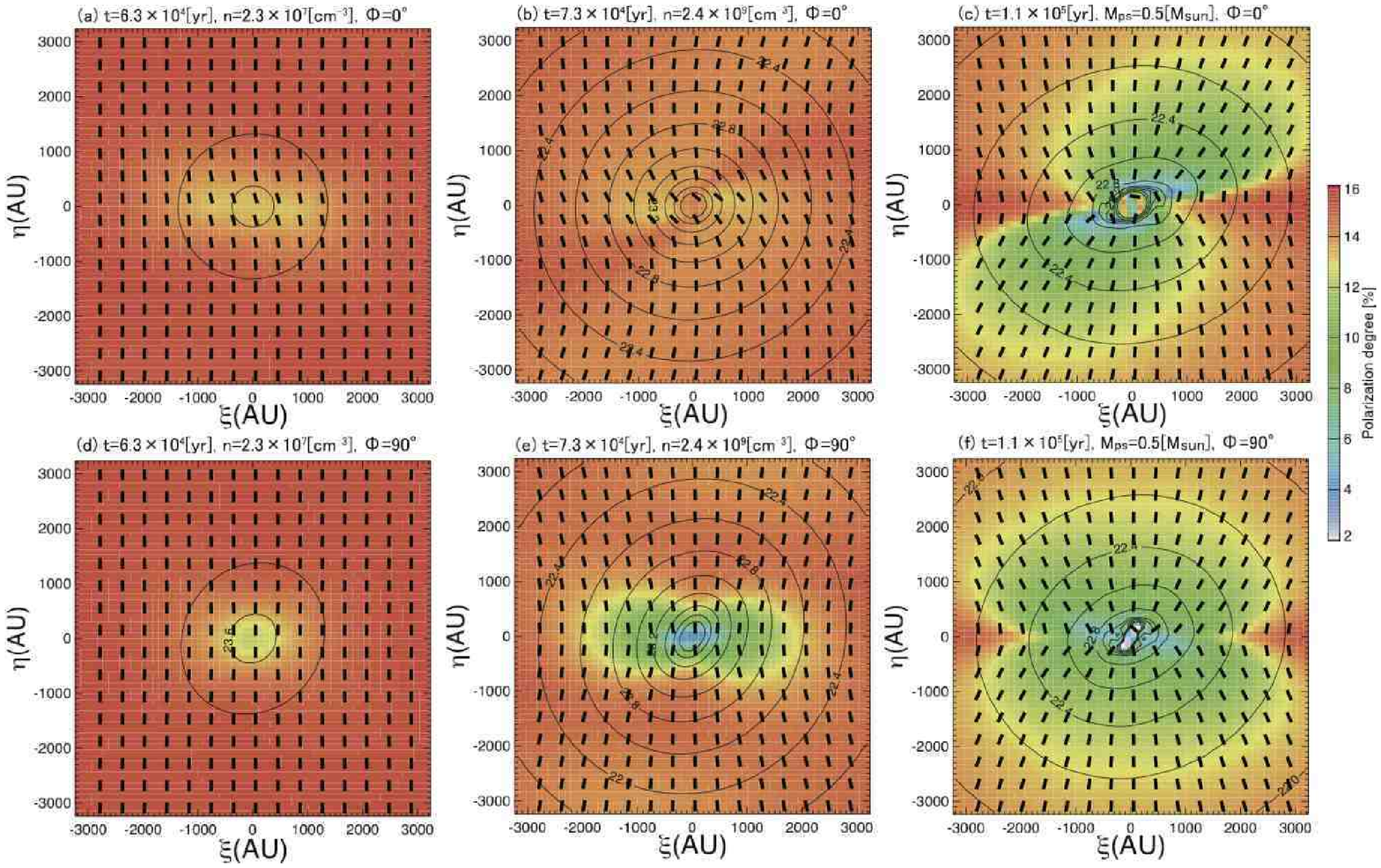}
\end{center}
\caption{
Polarization maps for model 3 at different epochs viewed from 
 ($\theta$, $\phi$) = ($90\degr$, 0$\degr$) (a, b, and c)
 and ($\theta$, $\phi$) = ($90\degr$,90$\degr$) (d, e, and f).
Panels (a) and (d) ($t=6.3\times 10^4 {\rm yr}$)
 and panels (b) and (e) ($t=7.3\times 10^4 {\rm yr}$) correspond to
 the prestellar contraction phase.
In contrast, panels (c) and (f) cover the protostellar phase
 with a protostar mass of $0.5\msun$ ($t=1.1\times 10^5 {\rm yr}$).
}
\label{fig:weakB_time}
\end{figure*}

In \S\ref{sec:rotation}, we investigated the polarization distribution in a cloud with an idealized initial setting in which the rotation axis is parallel to the magnetic field lines.
In reality, however, the rotation axis is not necessarily parallel to the magnetic field lines in molecular cloud cores.
In this subsection, we investigate the polarization in model 3 in which the initial cloud has a rotation axis that is not parallel to the magnetic field lines (see Figure~\ref{fig:initial}). 
In model 3, the initial angle between the rotation axis and the magnetic field lines is $\ang=60\degr$, and the rotational energy ($\beta_0=0.02$) is equivalent to the magnetic energy ($\gamma_0=0.02$) in the initial cloud.
Figure~\ref{fig:weakB_time} shows polarization maps at two viewing angles, $(\theta,\phi)= (90\degr,0\degr)$ ([a], [b], and [c]) and $(\theta,\phi)= (90\degr,90\degr)$ ([d], [e], and [f]), for model 3 at three different epochs, in which two snapshots ($t=6.3\times 10^4 {\rm yr}$ and $t=7.3\times 10^4 {\rm yr}$) describe the prestellar phase ([a], [b], [d], and [e]) and one snapshot ($t=1.1\times 10^5 {\rm yr}$) represents the protostellar phase ([c] and [f]).
At an age of $t=1.1\times 10^5 {\rm yr}$, the protostar has grown to $M_{\rm ps}=0.5 \msun$.
As in equation~(\ref{eq:omg-angle}), the rotation axis is inclined by $\ang=60\degr$ from the $z$-axis on the $xz$ plane in the simulation grid.
Because the $yz$ plane in the simulation grid coincides with the $\xi \eta$ plane in the observation grid when viewed from $(\theta,\phi)= (90\degr,0\degr)$, the initial rotation axis is inclined by $\ang=60\degr$ from the $\eta$-axis to the observer in the upper panels of Figure~\ref{fig:weakB_time} (see Figure~\ref{fig:grid}).
In the lower panels, where $(\theta,\phi)= (90\degr,90\degr)$, the initial angular momentum vector is directed toward  $(\xi,\eta)=(-\sqrt{3}/2,+1/2)$. 

\begin{figure}
\begin{center}
\includegraphics[width= 80mm]{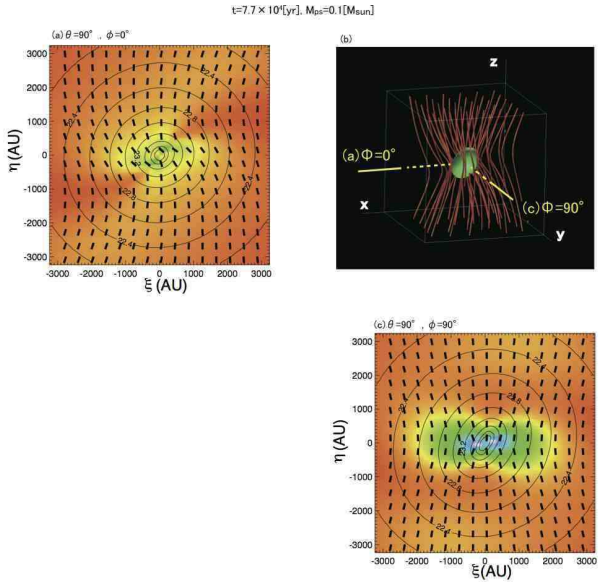}
\end{center}

\caption{
Polarization maps when $M_{\rm ps} =0.1\msun$ (a and c) for model 3.
Viewing angles are $(\theta,\phi)=(90\degr,0\degr)$ (a)
 and $(\theta,\phi)=(90\degr,90\degr)$ (c).
Panel (b) represents the three-dimensional distribution of the magnetic field lines
 and isodensity surface of $n=4.7 \times 10^{5}\cm$.
}
\label{fig:weakB_phi}
\end{figure}

Figure~\ref{fig:weakB_time} shows that the polarization degree around the center of the cloud decreases with time because the low-polarization region develops after the protostellar grows sufficiently ([c] and [f], in which $M_{\rm ps}=0.5 \msun$).
The reduction rate in the polarization degree for model 3 (Figure~\ref{fig:weakB_time}) is similar to that for model 2 (Figure~\ref{fig:rotation_time}).
However, a line symmetry with respect to the $\eta$-axis is broken in model 3 (panel [b]) but maintained in model 2 (Figure~\ref{fig:rotation_time}).
In addition, the polarization vectors in Figures~\ref{fig:weakB_time}(b) and (e) differ considerably from those in Figures~\ref{fig:standard_time} and \ref{fig:rotation_time}.
For models 1 and 2, the magnetic field lines have an hourglass configuration for any viewing angle $\phi$ (and $\theta$), indicating that they have axisymmetry around the $z$-axis.
On the other hand, for model 3, the polarization vectors appear different at viewing angles of $\phi=0\degr$ (b) and $\phi=90\degr$ (e).
Panels (a) and (d), which represent the structure before core formation, indicate a slight asymmetry with respect to the $\eta$-axis at both viewing angles.
This means that in the prestellar phase, the departure from axisymmetry around the $z$-axis is weak.
Panels (c) and (f), which show polarization maps of the protostar phase with $M_{\rm ps}=0.5\msun$, seem to indicate that the axisymmetry in the polarization vectors is recovered again.
Therefore, the departure from the axisymmetric structure is prominent in the final phase of prestellar contraction (panels [b] and [e]) and the early protostellar phase (Figure~\ref{fig:weakB_phi}).
This non-axisymmetric magnetic field has the shape of the letter $S$ in panel (b).
That is, in panel (b), the polarization vectors change their directions from the vertical direction in the uppermost region to the direction from the upper left to the lower right around the equator ($\eta=0$), and finally to the vertical direction in the lowermost region.
The non-axisymmetric magnetic field appears in another way in panel (e): the major axis of the total intensity distribution (column density) extends from the upper right to the lower left.
Although the polarization vector is essentially perpendicular to the disk major axis in models 1 and 2, the polarization vector in panel (e) indicates an hourglass shape, but its axis is not perpendicular to the major axis of the disk. 

As described in \S\ref{sec:rotation}, the polarization vectors do not depend on the azimuthal viewing angle $\phi$ in a cloud with an initial rotation axis parallel to magnetic field lines.
On the other hand, they do depend on $\phi$ when the initial magnetic field lines are misaligned with the rotation axis (Figure~\ref{fig:weakB_time}).
Figure~\ref{fig:weakB_phi} shows polarization maps for $\theta=90\degr$ for model 3 at the epoch when the protostar has a mass of $M_{\rm ps}=0.1\msun$ and a three-dimensional view of the magnetic field and isodensity surface (panel [b]).
The viewing angle $\phi$ is equal to $0\degr$ in panel (a) and $90\degr$ in panel (c).
The protostellar mass of $M_{\rm ps}=0.1\msun$ is smaller than $M_{\rm ps}=0.5\msun$ in Figures~\ref{fig:weakB_time}(c) and (f).
This clearly shows that in the early evolutionary stage of protostar accretion, the viewing angle has a significant effect.
Note that the early protostellar phase (Figures~\ref{fig:weakB_phi}[a] and [c]) is very similar to the final phase of the prestellar stage (Figures~\ref{fig:weakB_time}[b] and [e]).
The polarization vectors have the $S$-shaped configuration in Figure~\ref{fig:weakB_phi}(a) at $\phi=0\degr$, whereas they have an hourglass configuration in Figure~\ref{fig:weakB_phi}(c) at $\phi=90\degr$.
Although the disk is inclined with respect to the $z$-axis (the isodensity surface in Figure~\ref{fig:weakB_phi}[b] and the surface density contours in Figure~\ref{fig:weakB_phi}[c]), the axis of the hourglass-shaped magnetic field seems parallel to the $z$-axis. 
In addition, each panel shows a different distribution of the polarization degree.
A low-polarization region appears in the lower-right and upper-left regions toward the protostar for $\phi=0\degr$ (panel [a]), whereas it appears almost along the $\xi$-axis near the protostar for $\phi=90\degr$ (panel [c]).
In reality, for model 3, the magnetic field lines do not show a clear hourglass structure in three dimensions.
Nevertheless, depending on the viewing angle, they appear as an hourglass shape, as shown in Figure~\ref{fig:weakB_phi}(c). 

The distribution of the polarization degree and polarization vectors for $\phi=90\degr$ seems to be almost the same as that in the aligned rotation case (model 2).
However, the column density contour in Figure~\ref{fig:weakB_phi}(c) indicates that a pseudo-disk is inclined by $\sim45\degr$ from the $\xi$-axis.
This is an outcome of the inclined density distribution shown in Figure~\ref{fig:weakB_phi}(b).
That is, when viewed from the $y$-axis, the disk (isodensity surface) also extends from the lower left to the upper right.
On the other hand, the column density has a spherical symmetry in Figure~\ref{fig:weakB_phi}(a) for $\phi=0\degr$.
Figure~\ref{fig:weakB_phi}(b) shows that the symmetry rotation axis of the disk is in the $xz$ plane, and the disk normal is directed to approximately $\sim45\degr$ from the $x$-axis. 

The deviation from the hourglass configuration occurs because the angular momentum is not parallel to the global magnetic field.
However, as the cloud collapses, the angular momentum perpendicular to the global magnetic field lines is effectively transferred by magnetic braking \citep{price07}.
Thus, at a later evolutionary stage, the rotation axis tends to align with the global magnetic field lines.
Therefore, the rotation axis becomes parallel to the magnetic field, and the hourglass configuration is recovered.
However, because the angular momentum of the rotating disk is not completely transferred, the disk normal is not completely aligned with the global magnetic field, especially at the small scale ($\sim500$ AU), as seen in the central regions of Figures~\ref{fig:weakB_time}(c) and (f).

Cloud rotation has a significant effect on the star formation process because it produces a circumstellar disk where planets form.
It is thought that magnetic field lines are generally not well aligned with the rotation axis in molecular cloud cores. 
Thus, we expect that the $S$-shaped configuration of the magnetic field lines seen in Figures~\ref{fig:weakB_time} and ~\ref{fig:weakB_phi} will be frequently observed in the early stage of star formation in future high-angular-resolution observations.

\subsubsection{Model 4: Misaligned Cloud with Strong Magnetic Field}
As described in \S\ref{sec:weakB}, when the rotation axis is inclined from the magnetic field lines, the polarization vectors do not indicate hourglass-shaped but $S$-shaped structure.
The $S$-shaped configuration is caused by cloud rotation, which can cause the magnetic field lines to deviate from the hourglass configuration.
Both the magnetic field (Lorentz force) and rotation (centrifugal force) can form a disk in the collapsing cloud.
When a gas cloud having a strong magnetic field collapses along the magnetic field lines, a pseudo-disk forms in the direction perpendicular to the magnetic field lines.
On the other hand, when a gas cloud having a larger angular momentum collapses along the rotation axis, a rotating disk is extended in the direction perpendicular to the rotation axis.
Thus, it is expected that the former creates a clear hourglass polarization configuration and the latter shows the $S$-shaped configuration, depending on the viewing angle, when the magnetic field lines are not parallel to the rotation axis.
We confirmed the $S$-shaped configuration with a relatively weak magnetic field in \S\ref{sec:weakB} (i.e., the former case).
In this subsection, to investigate the latter case, we show polarization maps of model 4.
Model 4 initially has a strong magnetic field ($\gamma_0=0.57$) but weak rotation ($\beta_0=0.01$).

\begin{figure}
\begin{center}
\includegraphics[width= 80mm]{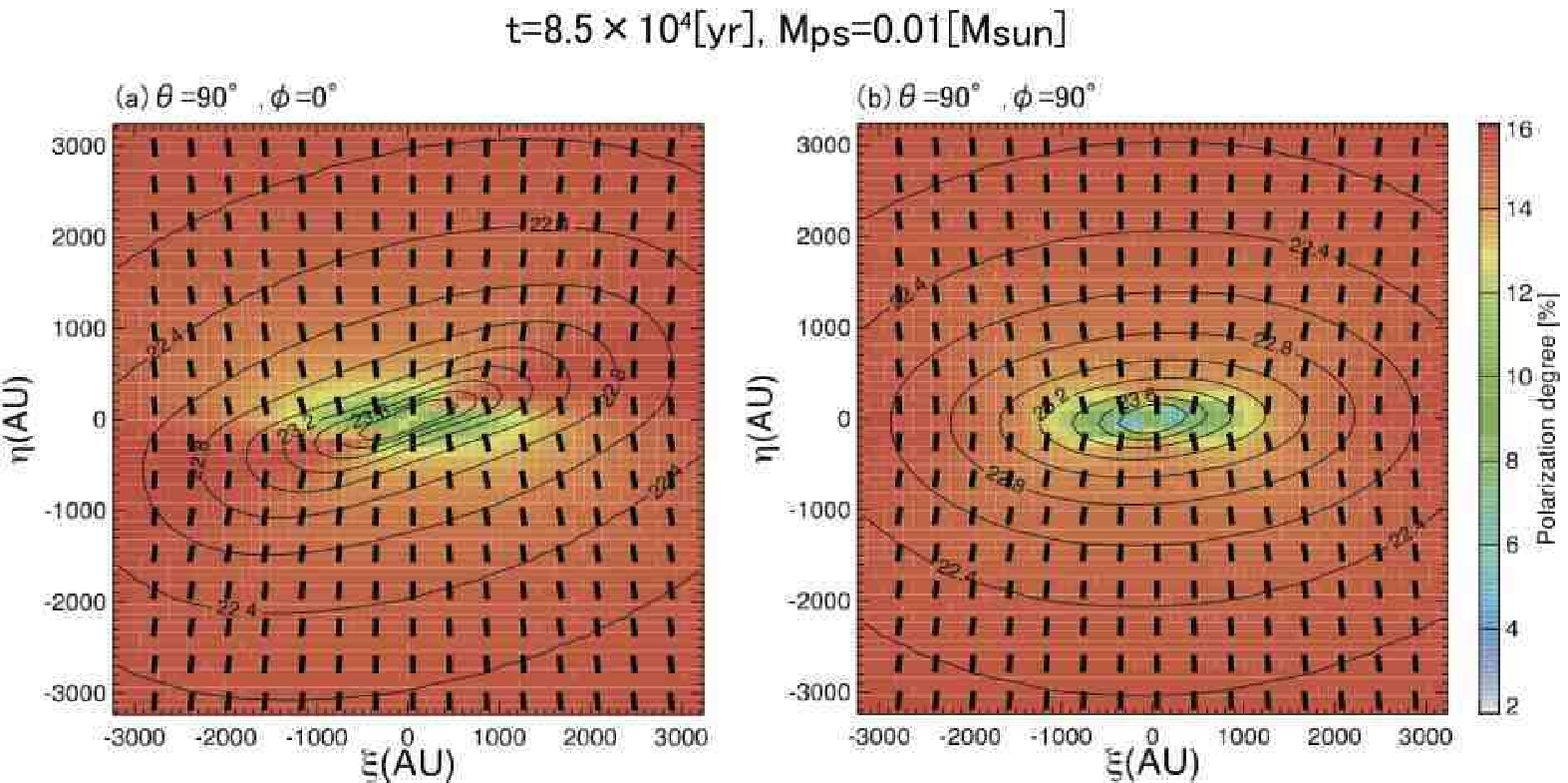}
\end{center}
\caption{
Polarization maps for model 4 when $M_{\rm ps} = 0.01\msun$
 at viewing angles of $(\theta,\phi)=(90\degr,0\degr)$ (a)
 and $(\theta,\phi)=(90\degr,90\degr)$ (b).
}
\label{fig:strongB_phi}
\end{figure}

Figure~\ref{fig:strongB_phi} shows polarization maps of model 4 viewed from the angles of $(\theta,\phi)=(90\degr,0\degr)$ (a) and $(\theta,\phi)=(90\degr,90\degr)$ (b) just after the protostar formation epoch ($M_{\rm ps} = 0.01\msun$). 
The figure indicates that polarization vectors have the hourglass configuration irrespective of the viewing angle $\phi$. 
On the other hand, in model 3, the polarization vectors exhibit the $S$-shaped configuration even when $M_{\rm ps} = 0.1\msun$.
We confirmed that for model 4, the hourglass configuration is maintained until a later evolutionary stage.
Thus, we conclude that in a cloud with a relatively strong magnetic field, rotation rarely causes the magnetic field lines to deviate from the hourglass configuration.
This is because gas falls onto the central region mainly along the magnetic field lines, not along the rotation axis.

At a small scale, however, the line symmetry of the polarization vectors with respect to the $\xi$-axis seems to be broken.
In addition, the column density contours in Figure~\ref{fig:strongB_phi} imply an inclined disk.
Thus, the effect of the misalignment between the magnetic field and the rotation axis cannot be negligible even for this model.
The deviation of the magnetic field lines from the hourglass configuration very close to the protostar may be detectable with future instruments such as ALMA.

\section{Discussion}

\subsection{Polarization Angle and Configuration of Magnetic Field Lines}
\label{sec:angle}
To date, magnetic field lines are generally considered to have an hourglass structure in a gravitationally contracting cloud.
However, as described in \S\ref{sec:visualization}, we showed that this hourglass configuration is not always realized in a collapsing cloud.
Depending on the viewing angle, the magnetic field lines have an $S$-shaped configuration when the cloud rotation influences the dynamical cloud evolution.
In any case (hourglass or $S$-shaped configurations), the magnetic field lines gradually deviate from their initial configuration as the cloud collapses.
In this subsection, we present a method of measuring the magnetic field structure in star-forming clouds by qualitatively estimating the magnetic field line structure.

\begin{figure}
\begin{center}
\includegraphics[width= 80mm]{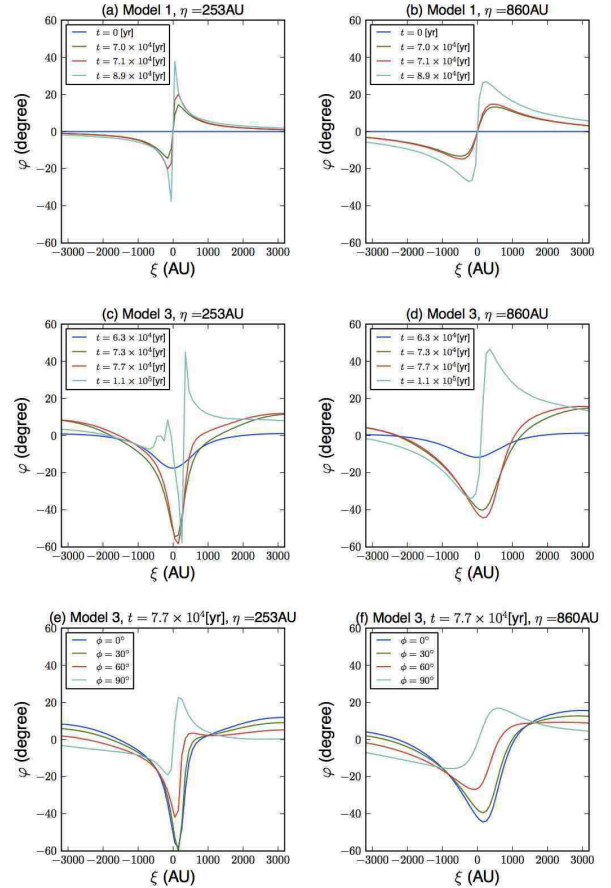}
\end{center}
\caption{
Time sequence of deviation angle of polarization vectors
 from the initial configuration for models 1 (a and b) and 3 (c and d).
Deviation angle was calculated along the $\xi$-axis at $\eta=253$\,AU for (a) and (c)
 and at $\eta=860$\,AU for (b) and (d) on the polarization map with
 ($\theta$, $\phi$) = ($90\degr$, $0\degr$).
Viewing angle dependence is shown in panels (e) and (f),
 in which snapshots of model 3 at $t=7.7\times 10^4 {\rm yr}$ are shown for
 $\phi=0\degr$, $30\degr$, $60\degr$, and $90\degr$.  
}
\label{fig:hourglass_standard}
\end{figure}

In the low-density gas region ($n<10^{11}\cm$), the magnetic field is well coupled with neutrals via momentum exchange with ions.
Thus, as the cloud collapses, the magnetic field lines converge toward the center of the cloud and form the hourglass structure, when they are initially parallel to the rotation axis.
If we simply assume an initially uniform magnetic field, the transverse motion of the gas drags the magnetic field lines.
In a cloud, the gas initially distributed at a distance from the cloud center must move a long distance and later fall onto the cloud center (or onto the protostar or circumstellar disk).
Thus, the deviation of the magnetic field from the initial configuration is believed to propagate from the center to the outer envelope with time.
To qualitatively investigate this, we calculated the deviation angle of the polarization vectors from the initial configuration on the polarization map with ($\theta, \phi)= (90\degr$, $0\degr$)
\footnote{We owe the idea of figure~\ref{fig:hourglass_standard} to S.Basu, who presented similar diagrams in Winter School and Workshop on Star Formation in Tokyo, 2011 (see also \citet{Basu09a, Basu09b})}.
Figure~\ref{fig:hourglass_standard} shows a time sequence of the deviation angle in which the horizontal axis coincides with the $\xi$-axis.
The angle is set to $0\degr$ for the initial magnetic field lines.
The deviation angle is measured along a straight line parallel to the $\xi$-axis at $\eta=253$\,AU in panels (a), (c), and (e) and $860$\,AU in panels (b), (d), and (f).

Figures~\ref{fig:hourglass_standard}(a) and (b) show the deviation angle for model 1.
Model 1 has no rotation and thus exhibits a clear hourglass, as shown in Figure~\ref{fig:standard_time}.
The snapshots are the same in Figure \ref{fig:standard_time} and Figures~\ref{fig:hourglass_standard}(a) and (b).
Figure~\ref{fig:hourglass_standard}(a) shows that the deviation angle has two sharp extremes at $r\sim \pm (100-200)$\,AU.
The maximum deviation angle is $\sim10\degr$ before protostar formation and $\lesssim40 \degr$ after it.
That is, the deviation angle gradually increases overall.
This indicates that the magnetic field lines are gradually inclined from their initial configuration.
In addition, the deviation angle has a point symmetry with respect to the origin.
The point symmetry in this plot is an indicator of the hourglass configuration of the magnetic field lines.
Figure~\ref{fig:hourglass_standard}(a) shows that the deviation angle at a larger distance ($\gtrsim 1000-2000$\,AU) remains $\lesssim5 \degr$.
This is because the deviation angle in this panel is measured along the line near the equator.
As shown in Figure~\ref{fig:standard_time}, the polarization vectors are inclined very little near the $\eta=0$ symmetry axis; this is because the magnetic field lines on the equatorial plane are pulled mainly in the direction perpendicular to their initial direction (or in the direction toward the protostar).

A comparison of panels (a) and (b) indicates that the absolute values of the extremes decrease as the height from the mid-plane increases.
The reason is that the magnetic field is dragged by the gas, which is contracting toward the center.
However, at a distance from the center in the $\xi$ direction, the deviation angle is larger for a larger height (panel [b]).
This seems to be explained by the fact that because of the mirror symmetry with respect to the $z=0$ plane, the magnetic field near the $z=0$ plane must be aligned along the $z$-axis.
Away from the $z=0$ plane, the magnetic field is bent more strongly.

Figures~\ref{fig:hourglass_standard}(c) and (d), in which the snapshots are the same as in Figures \ref{fig:weakB_time} and \ref{fig:weakB_phi}, show that the point symmetry is broken in model 3.
In model 3, the initial rotation axis is inclined from the global magnetic field lines, and the $S$-shaped configuration of the polarization vectors appears ($t=7.3\times 10^4 {\rm yr}$ in Figure~\ref{fig:weakB_time}[b] and $t=7.7\times 10^4 {\rm yr}$ in Figure\ref{fig:weakB_phi}).
In the panel, a single negative extreme appears at $\xi \sim 200-300$\,AU in the early evolutionary stage, which represents the $S$-shaped configuration of the polarization vectors (Figure~\ref{fig:weakB_time}[b]).
On the other hand, another positive extreme develops gradually with time, and the deviation angle begins to have two extremes (one positive and one negative).
This means that the magnetic field configuration changes from the $S$-shaped to the hourglass with time.
This deviation angle plot can clearly distinguish between the hourglass and $S$-shaped configurations.

Figures~\ref{fig:hourglass_standard}(e) and (f) show the viewing angle dependence of $\phi$.
Viewed from $\phi=90\degr$, the deviation angle of model 3 has two extremes, one positive and another negative.
However, the same model shows a single extreme if we observe from $\phi\le 60\degr$.
This clearly shows that in these periods the $S$-shaped configuration ($\phi\lesssim 60\degr$) and the hourglass configuration ($\phi\simeq 90\degr$) are realized simultaneously and that the appearance depends on the viewing angle $\phi$.

\subsection{Toroidal Components on Polarization Map}
In a non-rotating collapsing cloud, only the poloidal component of the magnetic field is generated.
In contrast, the toroidal field appears in addition to the poloidal field in rotating clouds.
The protostellar outflow and angular momentum transfer due to magnetic braking are closely related to the cloud rotation and toroidal field \citep{tomisaka02,machida11}.
Thus, the toroidal field is crucial to identify the driving mechanism of protostellar outflows or investigate the transfer mechanism of angular momentum.
In this subsection, we present the observation of toroidal components on the polarization maps.

When we observe the star-forming cloud from the pole (pole-on view), we can directly observe the toroidal field.
We can confirm the presence of the toroidal component of the polarization vectors in Figure~\ref{fig:rotation_theta}(a) (pole-on view of polarization map). 
However, the polarization degree is very low at the large (or cloud) scale because the poloidal component dominates the toroidal component.
Note that because the freefall timescale is shorter than the rotation timescale at the cloud scale, the toroidal component develops very little in such a region.
Although the toroidal field is generated around the protostar by disk rotation, the polarization degree is as low as $\lesssim$5\% at the cloud scale. 
Thus, it is difficult to determine the toroidal fields from such observations.

Another way to detect the toroidal field is to investigate the direction of the polarization vectors on the polarization map with a viewing angle of $\theta>0\degr$.
That is, as shown in the bottom panels of Figure~\ref{fig:rotation_theta}, the polarization vectors just above and below the protostar ($r<500$\,AU) do not point to the cloud center; they are almost perpendicular to the global field lines andparallel to the $\xi$-axis.
In addition, the polarization degree in this region is extremely low.
This is because the foreground and background dust alignments cancel each other.
Although it is difficult to determine the polarization direction in such a low-polarization region, the existence of the region of extremely low polarization above and below the protostar is an indicator of the toroidal field. 

The toroidal field can also be confirmed by observing a point symmetry of the low-polarization region around the protostar.
As described in \S\ref{sec:weakB} and \citet{tomisaka11}, both the viewing angle effect and the toroidal component of the magnetic field produce a point-symmetric distribution of a low polarization degree with respect to the protostar on the polarization map (compare Figures~\ref{fig:standard_theta} and \ref{fig:rotation_theta}).
The point-symmetric polarization distribution implies the existence of a toroidal field. 
Although we cannot simply determine the existence of the toroidal field, we can confirm it by combining evidence appearing in the polarization map, such as the direction of the polarization vectors and the distribution and symmetry of the polarization degree.

\subsection{Comparison with Observations}
\begin{figure}
\begin{center}
\includegraphics[width= 80mm]{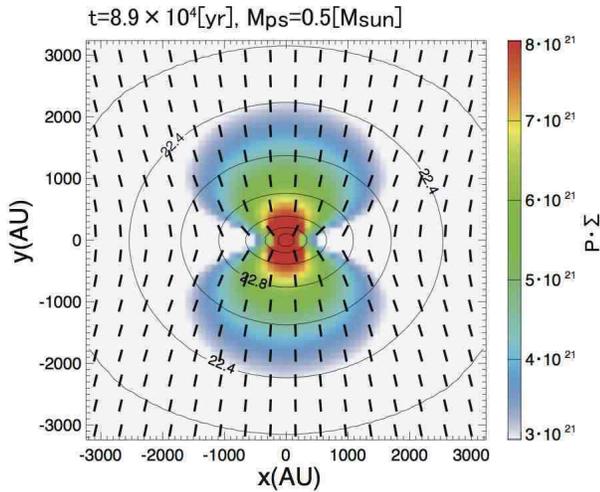}
\end{center}

\caption{
Distributions of polarization degree times column density (colors), which is proportional to polarized intensity; polarization vector (lines); and column density (contours, in $\,{\rm cm}^{-2}$ ) for model 1 at a viewing angle of ($\theta$, $\phi$) = ($60\degr$, $0\degr$).
The protostellar mass is $M_{\rm ps}=0.5\msun$ at this epoch.
}
\label{fig:compare_observation_girart}
\end{figure}

In this subsection, we compare our results with observations.
We can create a polarization map from simulation data at an arbitrary viewing angle to reproduce polarization observations. 
First, we compare our results with the observations of \citet{girart09}.
Using the SMA, \citet{girart09} observed a molecular cloud core in star-forming region G31.41+0.31 and showed the polarization distribution.
To fit the polarization observation shown in \citet{girart09}, we chose model 1 that shows a clear hourglass structure in three dimensions.
Figure~\ref{fig:compare_observation_girart} shows the polarization degree times the column density (colors), which is proportional to the polarized intensity; the polarization vector (lines); and the column density (contours) for a viewing angle of ($\theta$, $\phi$) = ($60\degr$, $0\degr$).
The left panel of Figure~1 of \citet{girart09} shows two peaks of polarized intensity above and below the protostar. 
In addition, a low-polarization region appears next to the protostar in the direction perpendicular to the expected global magnetic lines.
These features are reproduced well in Figure~\ref{fig:compare_observation_girart}.
Note that these features do not clearly appear at a viewing angle of $\theta=0\degr$, as shown in Figure~\ref{fig:standard_theta}.
As described in \S\ref{sec:standard}, these features (two peaks and low-polarization regions) are emphasized as the viewing angle $\theta$ increases.
Thus, it is believed that this star-forming core was observed with $\theta\ne0$ with respect to the global field lines.
Therefore, this structure is very sensitive to the inclination angle.
In addition, we can imagine not a point symmetry but a line symmetry of polarization intensity in Figure~1 of \citet{girart09}.
This indicates that a magnetic field, not rotation, controls the cloud evolution because a point symmetry appears when the rotation energy dominates the magnetic energy (see Figure~\ref{fig:weakB_time}).
A line symmetry is also confirmed in Figure~\ref{fig:compare_observation_girart}.

Next, we focus on the polarization observation of \citet{shinnaga12}.
They measured polarized dust emission toward the high-mass star-forming clump IRAS 20126+4104.
In contrast to the star-forming cloud observed by \citet{girart09}, the cloud rotation seems to significantly affect the dynamical cloud evolution.
\citet{shinnaga12} showed the $S$-shaped distribution of the polarization vectors, which indicates that (i) the magnetic field lines are not aligned with the rotation axis and (ii) the rotational energy is larger than or comparable to the magnetic energy in the cloud.
Fortunately, the star-forming clump IRAS 20126+4104 was observed at various wavelengths, and the directions of the high-speed jet, molecular outflow, and circumstellar disk were determined \citep{shinnaga12}.
In addition, the rotation axis of the entire cloud and the global magnetic field direction were also observed.
The observations indicate that the cloud rotation axis is inclined from the global magnetic field in the star-forming clump IRAS 20126+4104.
In such a case, the magnetic field lines deviate from an hourglass configuration and have an $S$-shaped configuration when the rotational energy dominates the magnetic energy, as described in \S\ref{sec:weakB}.
The $S$-shaped configuration is clearly seen in Figure 1 of \citet{shinnaga12}.
Thus, their polarization observations agree well with our theoretical prediction.

Structure of magnetic field different from hourglass shape is also observed in a low-mass forming region IRAS $16293-2422$ \citep{rao09}.
They observed the object with the SMA in the 345 GHz band and obtained maps with molecular lines such as ${\rm H^{13}CO^+}\ J=4-3$, ${\rm SiO}\ J=8-7$ and ${\rm CO}\ J=3-2$ in addition to the dust continuum polarization.
Using both polarization and line emission observations, they obtained an interesting configuration between the magnetic field and the rotation. 
That is, in IRAS $16293-2422$, the rotational axis seems perpendicular to the magnetic field around source A. 
This is similar to the $S$-type configuration observed in Figure~\ref{fig:weakB_phi}(c) of model 3, in which the polarization B-vector is likely to be perpendicular to the rotational axis in the very center (see also \S\ref{sec:velocity}). 
Therefore, we can explain the structure of magnetic field in this object partly by using S-type shape structure.
However, since this object is binary, the binary system may greatly affect the polarization patterns.
We postpone the polarization pattern expected for binary-forming model for a future separate paper.

It should be noted that we should be careful about the filtering effect on the large scales of the interferometers. 
However, this study simply focuses on classifying polarization patterns with MHD simulations. Therefore, we do not consider any effects of interferometers.
We should consider such effects for specific cases in future works.

\subsection{Effects of Sink Cells}
\label{sec:effect-of-sink}
To investigate the long-term evolution of the magnetic field in star-forming cloud cores, we adopted sink cells, as described in \S\ref{sec:sink}.
With sink cells, we did not resolve the high-density gas region around the protostar (or the center of the cloud). 
In the low-density gas region (or at large scale), the magnetic field is well coupled with neutrals and the magnetic field lines are accompanied by neutral gas motion.
In principle, the gas motion in the low-density region is affected very little by the inner high-density gas region because the gas continues to fall onto the center of the cloud, and the gas motion is determined by the gravity as well as the local magnetic field, rotation, and pressure gradient force.
Note that we correctly calculated the gravity inside the sink by adding the gas removed from inside the sink to the gravitational potential (see \S\ref{sec:sink}).
Thus, we believe that we do not always resolve the region very close to the protostar ($\ll10$\,AU) when we investigate the evolution of cloud-scale magnetic field lines ($\sim1000$\,AU).
In this study, although we adopted a sink radius of $r_{\rm sink}=15\,$AU, we could correctly investigate the magnetic field lines in the region of $r\gg15$\,AU.
In addition, in our previous studies \citep{machida12}, we confirmed that the magnetic field lines at the cloud scale changed very little even when we adopted a smaller sink radius.

The protostellar outflow may affect the large-scale magnetic field structure.
The molecular outflow is believed to be driven by the first core \citep{larson69,masunaga00} before protostar formation \citep{tomisaka02} and by the circumstellar disk after protostar formation \citep{machida12}.
In this study, because both the first core and the circumstellar disk are not resolved with sufficient resolution, no powerful outflow appears; only a weak outflow appears for model 2.
However, both observations and numerical simulations have shown that protostellar outflows are aligned with magnetic field lines. 
Because plasma moving along the magnetic field lines never bends the lines, the outflows are considered not to significantly change the magnetic field configuration.
Thus, we can safely ignore the effects of the protostellar outflow in studying the magnetic field configuration at the large scale.
As a first step, we believe that our simple setting adopted in this study helps us to understand the global evolution of magnetic field lines in the collapsing gas cloud.

In recent cloud-scale polarization observations, we could not clearly confirm the effect of protostellar outflow \citep{girart06,girart09,shinnaga12}. 
The reason is that the spatial resolution of such observations is not sufficient to resolve the effects of protostellar outflows.
When the protostellar outflow is driven by magnetocentrifugal or magnetic pressure gradient forces, the toroidal component of the magnetic field dominates the poloidal field inside the outflow.
In such a case, as shown in \citet{tomisaka11}, a very low polarization degree is realized inside the outflow because the toroidal components canceled out each other.
Thus, we must consider the effects of protostellar outflow in order to investigate the polarization in the star-forming core at a small scale.
To precisely investigate the effect of the magnetic configuration on the star formation process, we need a higher spatial resolution in both simulations and observations to resolve the protostellar outflow in future.

\subsection{Velocity Map}
\label{sec:velocity}
\begin{figure*}
\begin{center}
\includegraphics[width=120mm]{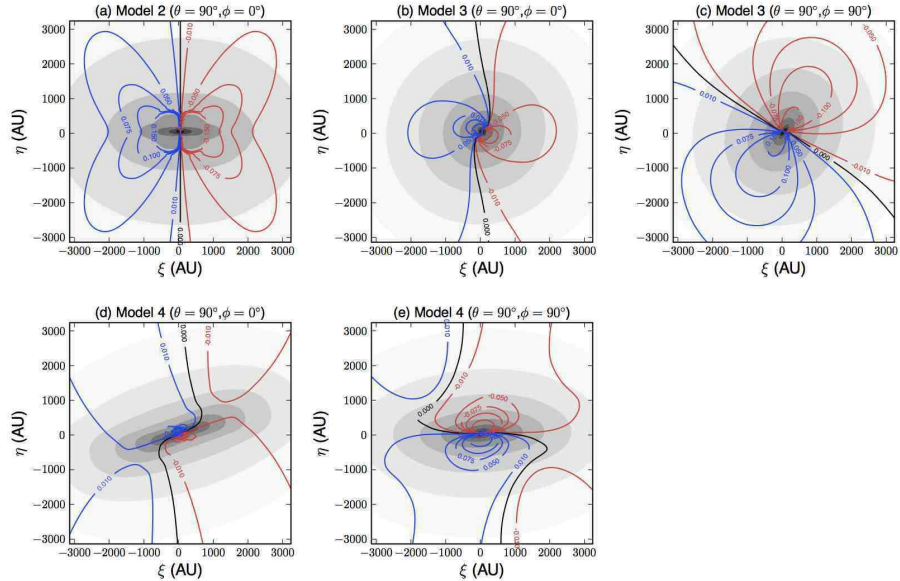}
\end{center}

\caption{
Velocity (red, blue and black contours) and column density (gray scale) maps for models 2, 3, and 4. 
In each panel, redshifted velocities of $\langle v \rangle = $ $-0.15$\,km\,s$^{-1}$, $-0.1$\,km\,s$^{-1}$, $-0.075$\,km\,s$^{-1}$, $-0.05$\,km\,s$^{-1}$ and $-0.01$\,km\,s$^{-1}$ are plotted with red contours while blueshifted velocities of $\langle v \rangle = $ 0.01\,km\,s$^{-1}$, 0.05\,km\,s$^{-1}$, 0.075\,km\,s$^{-1}$, 0.1\,km\,s$^{-1}$ and 0.15\,km\,s$^{-1}$ are plotted with blue contours.
The black contour corresponds to $\langle v \rangle = 0$.
Panels (a), (b), (c), (d) and (e) correspond to the polarization maps of Figures~\ref{fig:rotation_time}(c), \ref{fig:weakB_phi}(a), \ref{fig:weakB_phi}(c), \ref{fig:strongB_phi} (a) and \ref{fig:strongB_phi}(b), respectively.
}

\label{fig:velocity}
\end{figure*}

The gas motion is related to the configuration of magnetic field lines.
For example, the magnetic field lines (or the polarization vectors) are highly distorted by the cloud rotation as seen in Figures~\ref{fig:compare_rotation} and \ref{fig:weakB_phi}.
Thus, the velocity information is useful to understand the configuration and evolution of the magnetic field lines.
To investigate the relation between magnetic field lines and velocity fields, we calculated the first moment of velocity along the line-of-sight, 
\begin{equation}
\langle v\rangle = \dfrac{\int \rho\, {\bf  v} \cdot {\bf n}\, ds}{ \int \rho\, ds},
\label{eq:vr}
\end{equation}
where $ {\bf v} $  is velocity vector at each point and ${\bf n}$ is the unit vector along the line-of-sight.
With the average velocity [eq.~(\ref{eq:vr})], we made the velocity map for each model.
Figure~\ref{fig:velocity} shows the velocity distribution in the range of $ \vert \langle v \rangle \vert < 0.15\,{\rm km}$\,s$^{-1}$ for models 2, 3 and 4, in which each panel corresponds to the polarization map of Figures~\ref{fig:rotation_time}, \ref{fig:compare_rotation} and \ref{fig:weakB_phi}.
Note that the column density contour is superimposed for reference in Figure~\ref{fig:velocity}.

Figure~\ref{fig:velocity}(a) is  the edge-on view [($\theta$, $\phi$) = (90$\degr$, 0$\degr$)] of velocity map  for model 2, and corresponds to the polarization map of Figure~\ref{fig:rotation_time}(c).
In model 2, the initial rotation axis is parallel to the initial magnetic field lines.
From this panel, we can confirm a symmetric structure between blueshifted and redshifted velocity contours, indicating  that the rotation axis of infalling envelope and disk is parallel to the global magnetic field lines (or $\eta$- and $z$-axis).
Figures~\ref{fig:velocity}(b) and (c) are the velocity maps for model 3, and correspond to the polarization maps of Figures~\ref{fig:weakB_phi}(a) and (c), in which the viewing angles are  ($\theta$, $\phi$) = (90$\degr$, 0$\degr$) for Figure~\ref{fig:velocity}(b) and (90$\degr$, 90$\degr$) for Figure~\ref{fig:velocity}(c).
As described in \S4.2.3, for model 3, the initial rotation axis is inclined from the initial magnetic field lines, and thus the disk normal is not parallel to the magnetic field lines [see Fig.~\ref{fig:weakB_phi}(b)].
In Figures~\ref{fig:velocity}(b) and (c), the black contour indicating $\langle v \rangle=0$ near the center of the cloud ($r \lesssim$ 1000\,AU) roughly corresponds to the projected disk rotation axis  because the disk rotation velocity dominates the infall velocity in this region.
In Figure~\ref{fig:velocity}(b), the disk is tilted as the near side of the disk is lower and the far side is upper with respect to the line-of-sight.
On the other hand, we see the disk from the edge in Figure~\ref{fig:velocity}(c).
Thus, the rotation pattern in Figure~\ref{fig:velocity}(c) is more clear than that in Figure~\ref{fig:velocity}(b).
Comparison of Figure~\ref{fig:weakB_phi}(b) with Figures~\ref{fig:velocity}(b) and (c) indicates that the disk rotation axis is parallel to the disk normal.
Thus, we can identify the disk normal direction also from the velocity information.
As described in \S4.2.3, the magnetic field lines deviate from the hourglass configuration when the disk normal is not parallel to the magnetic field lines.

In contrast to models 2 and 3, the infall motion is more emphasized on the velocity map for model 4.
Figures~\ref{fig:velocity}(d) and (e) are the velocity maps for model 4, and correspond to the polarization maps of Figures~\ref{fig:strongB_phi}(a) and (b), in which the viewing angles are  ($\theta$, $\phi$) = (90$\degr$, 0$\degr$) for Figure~\ref{fig:velocity}(d) and (90$\degr$, 90$\degr$) for Figure~\ref{fig:velocity}(e).
For model 4, the initial rotation axis is inclined from the initial magnetic field lines as in model 3, while the initial strength of magnetic field is stronger than that for model 3.
In Figures~\ref{fig:velocity}(d) and (e), the column density distribution implies that the disk normal is roughly parallel to $\eta$-axis (or $z$-axis).
On the other hand, the velocity contour of $\langle v \rangle=0$ is almost perpendicular to the $\eta$-axis around the center of the cloud ($r \lesssim 1000$\,AU) and the velocity gradient is observed in the $\eta$-direction.
For model 4, a strong magnetic field effectively transfers the angular momentum by the magnetic braking mechanism as described in \S.4.1.
As a result, a tiny rotating disk is formed in the proximity of the protostar ($r\ll1000$\,AU).
Note that the disk-like structure of the column density contour in this panel corresponds to a pseudo-disk that is not supported by the rotation.
In addition, the rotation of the infalling envelope is also transferred by the magnetic braking.
Therefore, in these panels, the rotation motion is not apparent; the infall motion dominates the rotation motion especially in the scale of $\sim1000$\,AU.
In Figure~\ref{fig:velocity}(d), the disk is tilted as the near side is down and the far side is up.
Therefore, infalling gas along the disk is observed as a redshifted one in $\eta<0$ region but as a blueshifted gas in $\eta>0$ region. 
In contrast, the disk is tilted as the near side is up and the far side is down in Figure~\ref{fig:velocity}(e).
Thus, the infall motion is observed in an opposite way to (d).
Therefore, with the velocity information, we can understand the relation between the configuration of magnetic field and the disk or infalling envelope.

\section{Summary}
In this study, after we calculated the evolution of clouds with different initial parameters including magnetic field strength $B_0$, rotation rate $\Omega_0$, and angle $\delta_0$ between the magnetic field and the rotation axis, we created polarization maps from the data taken from three-dimensional MHD simulations.
Then, we investigated the polarization distribution and polarization vectors at different epochs with different viewing angles for each model and obtained the following results.

\begin{itemize}
\item In the prestellar stage, the polarization distributions of the models show no apparent features.
All the models exhibited a gradual decrease in the polarization degree around the center of the collapsing cloud. 
Strictly speaking, slight differences among the models appeared in the polarization distribution, depending on the initial cloud parameters of $\Omega_0$ and $\delta_0$ and the viewing angle.
However, we expect that such slight differences cannot be distinguished even in future polarization observations.
\item After protostar formation (i.e., in the protostellar phase), the polarization distribution showed some clear differences among the models.
In addition, it depended strongly on the viewing angle.
Depending on the viewing angle, a region of very low polarization appears next to the protostar because the radial component of the magnetic field around the protostar was canceled out.
Using the low-polarization region next to the protostar, we can estimate the angle between the direction of the global magnetic field and the line of sight.

\item Another difference is caused by the toroidal field.
In the protostellar phase, cloud rotation generates the toroidal field, which creates a region of extremely low polarization above and below the protostar.
On the other hand, no strong toroidal field appears in the prestellar stage because the freefall timescale is shorter than the rotation timescale.
Thus, we can predict the existence of the embedded protostar and circumstellar disk from such a region of extremely low polarization around the protostar.

\item The other difference is the symmetry of the polarization distribution.
Both the rotation and the toroidal field break a line symmetry and produce a point symmetry with respect to the position of the protostar.
The point-symmetric distribution of the polarization is an indicator of non-negligible rotation and a toroidal field.

\item The configuration of the polarization vectors depends strongly on the viewing angle in both the prestellar and protostellar stages.
We found that an hourglass-shaped magnetic field line structure is not always realized in the collapsing cloud.
Instead, an $S$-shaped configuration of polarization vectors (or magnetic field lines) often appears, depending on the viewing angle, when the global magnetic field lines are not aligned with the cloud rotation axis.
The $S$-shaped configuration is gradually transformed into the hourglass configuration.
Thus, we observed the $S$-shaped configuration of the magnetic field in the early evolutionary stage.

\item The $S$-shaped configuration appears in clouds having a rotational energy larger than the magnetic energy.
On the other hand, the hourglass configuration always appears independent of the viewing angle in clouds having a magnetic energy larger than the rotational energy.
Thus, from the configuration of the polarization vectors, we can predict the properties of the host cloud and the evolutionary stage of the protostar.

\end{itemize}

This study complements future polarization observations.
Without this type of effort, we cannot understand the real evolution and configuration of the magnetic field because the observed magnetic field is projected onto the celestial plane.
In this study, we showed the polarization distribution and polarization vectors while resolving considerably small-scale structures ($\sim10$\,AU).
Our results agree well with recent observations at the large (or cloud) scale, and we can determine their evolutionary stage and cloud properties.
We also created velocity maps to compare our results with observations, and showed that the combination of velocity and polarization information gives us a better understanding of the configuration of magnetic field lines and properties of the collapsing cloud core.
Although recent observations still lack sufficient spatial resolution to investigate small-scale structures around the protostar, our study is useful for understanding the effect of the magnetic field on the star formation process at a deeper level with near-future instruments such as ALMA.

\acknowledgments
We have greatly benefited from discussions with S. Shinnaga, S.H. Lai, and H.Nomura, which we acknowledge.
Numerical computations were performed on NEC SX-9 at the Center for Computational Astrophysics (CfCA) of the National Astronomical Observatory of Japan and on kashika at the Yukawa Institute Computer Facility.
This work was supported by Grants-in-Aid from MEXT [21244021 (KT) and 21740136 (MNM)].
We also thank to the anonymous referee for the helpful comments to improve the manuscript.
A.Kataoka is supported by the Research Fellowship from the Japan Society for the Promotion of Science (JSPS) for Young Scientists.

\end{document}